\documentclass[]{aa}
\usepackage[varg]{txfonts}
\usepackage{graphicx}
\usepackage{natbib}
\bibpunct{(}{)}{;}{a}{}{,} 

\title{Exploring the realm of scaled Solar System analogs with HARPS
	\thanks{Based on observations collected at the European Organisation for Astronomical Research in the Southern Hemisphere 
		under ESO programmes 093.C-0919(A) and 094.C-0901(A).}
		}
\subtitle{}
      
\author{D.~Barbato \inst{\ref{unito},\ref{oato}}
	\and A.~Sozzetti \inst{\ref{oato}}
	\and S.~Desidera \inst{\ref{oapd}}
	\and M.~Damasso \inst{\ref{oato}}
	\and A.S.~Bonomo \inst{\ref{oato}}
	\and P.~Giacobbe \inst{\ref{oato}}
	\and L.S.~Colombo \inst{\ref{unipd}}
	\and C.~Lazzoni \inst{\ref{unipd},\ref{oapd}}
	\and R.~Claudi \inst{\ref{oapd}}
	\and R.~Gratton \inst{\ref{oapd}}
	\and G.~LoCurto \inst{\ref{eso}}
	\and F.~Marzari \inst{\ref{unipd}}
	\and C.~Mordasini \inst{\ref{unibe}}
	}
\institute{Dipartimento di Fisica, Universit\`{a} degli Studi di Torino, via Pietro Giuria 1, I-10125 Torino, Italy \label{unito}
      \and INAF – Osservatorio Astrofisico di Torino, Via Osservatorio 20, I-10025 Pino Torinese, Italy \label{oato}
      \and INAF – Osservatorio Astronomico di Padova, Vicolo dell’Osservatorio 5, I-35122, Padova, Italy \label{oapd}
      \and Dipartimento di Fisica e Astronomia "G. Galilei", Universit\`{a} di Padova, Vicolo dell'Osservatorio 3, I-35122 Padova, Italy \label{unipd}
      \and European Southern Observatory, Alonso de Cordova 3107, Vitacura, Santiago, Chile \label{eso}
      \and Physikalisches Institut, University of Bern, Sidlerstrasse 5, 3012 Bern, Switzerland \label{unibe}
	}
	
\date{Received <date> / Accepted <date>}
\abstract{The assessment of the frequency of planetary systems reproducing the Solar System's architecture is still an open problem 
	  in exoplanetary science. Detailed study of multiplicity and architecture is generally hampered by limitations in quality, 
	  temporal extension and observing strategy, causing difficulties in detecting low-mass inner planets in the 
	  presence of outer giant planetary bodies.}
	  {We present the results of high-cadence and high-precision HARPS observations on 20 solar-type stars known to host a single 
	  long-period giant planet in order to search for additional inner companions and estimate the occurence rate $f_p$ of scaled 
	  Solar System analogs, i.e. systems featuring lower-mass inner planets in the presence of long-period giant planets.}
	  {We carry out combined fits of our HARPS data with literature radial velocities using differential evolution MCMC to refine 
	  the literature orbital solutions and search for additional inner planets. We then derive the survey detection limits to 
	  provide preliminary estimates of $f_p$.}
	  {We generally find better constrained orbital parameters for the known planets than those found in the literature; 
	  significant updates can be especially appreciated on half of the selected planetary systems. While no additional inner planet 
	  is detected, we find evidence for previously unreported long-period massive companions in systems HD 50499 and HD 73267. We 
	  finally estimate the frequency of inner low mass (10-30 M$_\oplus$) planets in the presence of outer giant planets as 
	  $f_p<9.84\%$ for P<150 days.}
	  {Our preliminary estimate of $f_p$ is significantly lower than the values found in the literature for similarly defined 
	  mass and period ranges; the 
	  lack of inner candidate planets found in our sample can also be seen as evidence corroborating the inward migration formation 
	  model for super-Earths and mini-Neptunes. Our results also underline the need for high-cadence and high-precision follow-up 
	  observations as the key to precisely determine the occurence of Solar System analogs.}

\begin{document}  
  \maketitle
  
  \section{Introduction} \label{sec:introduction}
    The science of exoplanetology has so far produced outstanding results: as of the time of writing (January 2018) the 
    NASA Exoplanet Archive contains data for 3587 confirmed exoplanets, 1509 of which are part of a total of 594 multiple planetary 
    systems. The recent increment in known multiple planets systems, arising in particular from the Kepler mission, has helped produce 
    the first comparative studies on their mass distribution and global architecture \citep[see][]{winn2015,hobson2017}, especially  
    focused on the multiplicity and architecture's role in supporting or disproving the current and competing formation models for both 
    giant and terrestrial planets as expanded upon in \citet{raymond2008}, \citet{cossou2014}, \citet{schlaufman2014} 
    and \citet{morbidelli2016}.
    \par A key parameter in discriminating between different formation models is the fraction of planetary systems featuring both gas 
    giants and lower-mass planets and their relative orbits; for example the numerical simulations reported 
    in \citet{izidoro2015} suggest that \textit{in situ} and \textit{inward migration} formation models of hot super-Earths, defined as 
    planets 1 to 20 times more massive than the Earth and with orbital periods lower than 100 days, would 
    cause very different systems architecture, the latter creating an anti-correlation between giant planets and close-in 
    super-Earth populations.
    \par However, the investigation of a particular characteristic of multiplanetary systems, namely the frequency of systems featuring 
    a clear mass and dimension hierarchy similar to the one dominating our Solar System, suffers from low-quality literature data and 
    would clearly benefit from larger statistics. Out of the 594 multiple planetary systems confirmed so far only two are generally 
    considered to be clear Solar System analogs, namely the GJ 676A \citep{sahlmann2016} and Kepler-90 \citep{cabrera2014} systems, and 
    these contain at least one super-Earth, a type of planet noticeably lacking in our own system.
    \par To further support this initial point, lacking a proper review on the subject, we have conducted a 
    preliminary search in the NASA Exoplanet Archive (as retrieved on September 12 2017) in order to provide the needed context on the 
    diverse known system architectures, focusing on those systems in which all planetary components have been studied via radial 
    velocity observations. Drawing the line between giant planets and smaller ones at 30 M$_\oplus$, during this search we found that 
    out of the 
    102 multiple planetary systems studied by RV surveys in which more than half of the planetary components have a catalogued 
    measure of mass, 57 host only giant planets, 24 host only lower-mass planets and only 11 
    (namely the systems around stars GJ 676 A, GJ 832, HD 10180, HD 11964, HD 164922, HD 181433, HD 190360, HD 204313, 
    HD 219134, HD 219828, HD 47186) are 
    what could be considered Solar System analogs, featuring inner low-mass planets and outer giant planets orbiting outside the 
    circumstellar habitable zone outer limit, computed here using the model discussed in \citet{kopparapu2013}. 
    Finally we note the existence of 10 mixed-architecture multiple systems, featuring giant planets on 
    close-in orbits or inside the habitable zone. We can therefore provide a preliminary estimate of known Solar System analogs 
    amounting to about 10\% of known multiple systems studied by RV surveys.
    \par The value of this fraction is obviously highly dependent on the exact definition of what constitutes a Solar System 
    analog. For example, it is significantly higher if one considers just the giant planets orbital periods instead of their position 
    relative to the system's habitable zone as key to architectural characterization; considering as analogs those systems composed of 
    inner lower-mass planets and outer giants whose orbits are longer than one year this fraction amounts to 15.57\% of known 
    multiple systems, and to 19.67\% if the lower limit on giant planets orbital period is instead set at 100 days.
    \par Although low, this fraction is an encouraging sign that a sizeable number of planetary systems may reproduce our own planetary 
    architecture, featuring at least one gas giant on a long-period orbit and lower-mass planets on inner orbits; 
    another positive clue can be identified in the findings of \citet{ciardi2013}, suggesting that in up to 60\% of Kepler candidate 
    planet pairs the innermost one has a smaller radius than the outer body, and that this fraction rises up to 68\% if the pair 
    contains one Neptune-sized planet. \citet{winn2015} also reports that approximately half of Sun-like star host at least one small 
    planet (from 1 to 4 R$_\oplus$) with an orbital period less than one year.
    \par During the aforementioned preliminary search through the RV-observed known systems we found 229 single-planet systems, 195 of 
    which contain a single long-period (P$\geq$1 yr) giant and 34 containing a single Neptune-like or super-terrestrial planet.
    In this context, it is clear that this large fraction of single known long-period giant planet systems may represent 
    an unparalleled opportunity for the search of yet undetected inner, smaller planets and therefore of scaled-down Solar System 
    analogs; however, known systems with giant planets are not usually subject to follow-up observations to search for additional, 
    lower 
    mass planetary companions at short orbital period. Furthermore, the available data published on most of these systems do not provide 
    by itself enough solid ground on which to base such a search, being characterized by a sampling frequency and in some case integration time 
    or instrument setup not suited for detecting low-mass and short period planets. To successfully detect such inner planets and their 
    very low amplitude signals compared to those of the more massive and distant companions, an intense monitoring of the host 
    star is needed in order to fully cover their orbital phase; the need for higher precision and more densely sampled observations on 
    already studied planetary systems is therefore clear and urgent in the search for scaled-down Solar System analogs.
    \par The purpose of this paper is therefore to present the results of the observations conducted by the HARPS spectrograph on a 
    small sample of these single-giant systems, in order to provide an estimate on the occurence rate of smaller 
    inner planets in the presence of a single outer giant planet and to fuel further observations; it is also of note that the analysis 
    of a parallel HARPS-N survey on a similarly selected northern hemisphere sample is currently underway and will be the subject of a 
    forthcoming paper.
    \par In Sect. \ref{sec:sample} we 
    describe the selection criteria for the analyzed sample, before moving on to describe the HARPS observations structure 
    and setup in Sect. \ref{sec:observations}; in Sect. \ref{sec:analysis} we show the results of the new data fits conducted to 
    refine the orbital parameters of the known planets. Finally, in Sect. \ref{sec:detectionlimits} we calculate the detection limits 
    for each system and the sample as a whole, in order to provide an estimate to the planetary frequency of low-mass inner planets in 
    the presence of outer giant planets to compare with the planetary frequency calculated in the seminal work of \citet{mayor2011} 
    for stars hosting at least one planet.
  
  \section{Sample selection and description}	\label{sec:sample}

    \begin{sidewaystable*}
      \small
      \caption{Stellar parameters for the sample systems}
      \label{table:samplestars}
      \centering
      \begin{tabular}{l c c c c c c c c c c c}
	\hline\hline
	Star & $\pi$ & V & B-V\tablefootmark{a} & T$_eff$ & log$g$\tablefootmark{a} & [Fe/H] & Mass\tablefootmark{a} & Radius\tablefootmark{a} & Age\tablefootmark{a} & $\log R^{\prime}_{HK}$ & P$_{rot}$\tablefootmark{e} \\
	     & [mas] & [mag] & [mag] & [K] & [cgs] & & [M$_\odot$] & [R$_\odot$] & [Gyr] &  & [d] \\
	\hline
	HD 4208   & $30.58\pm1.08$ & $7.795\pm0.011$ & $0.726\pm0.008$ & $5717\pm33$\tablefootmark{b} & $4.501\pm0.036$ & $-0.28\pm0.02$ & $0.883\pm0.024$ & $0.846\pm0.028$ & $3.813\pm2.970$  & -4.77  &   $\sim$25 \\[3pt]
	HD 23127  & $11.22\pm0.76$ & $8.576\pm0.002$ & $0.701\pm0.013$ & $5843\pm52$\tablefootmark{b} & $4.146\pm0.054$ & $0.29\pm0.03$  & $1.208\pm0.045$ & $1.490\pm0.104$ & $4.508\pm0.788$  & -5.00  &   $\sim$33 \\[3pt]
	HD 25171  & $17.84\pm0.60$ & $7.782\pm0.013$ & $0.618\pm0.005$ & $6125\pm21$\tablefootmark{b} & $4.263\pm0.031$ & $-0.12\pm0.04$ & $1.076\pm0.021$ & $1.230\pm0.045$ & $4.802\pm0.619$  & -4.99  &   $\sim$22 \\[3pt]
	HD 27631  & $22.45\pm0.78$ & $8.243\pm0.012$ & $0.721\pm0.009$ & $5737\pm36$ 		      & $4.455\pm0.038$ & $-0.12\pm0.05$ & $0.944\pm0.032$ & $0.923\pm0.033$ & $4.010\pm2.892$  & -4.91  &   $\sim$31 \\[3pt]
	HD 30177  & $19.93\pm0.63$ & $8.397$         & $0.773\pm0.012$ & $5607\pm47$\tablefootmark{b} & $4.417\pm0.034$ & $0.39\pm0.05$  & $1.053\pm0.023$ & $1.019\pm0.034$ & $2.525\pm1.954$  & -5.07  &   $\sim$45 \\[3pt]
	HD 38801  & $10.39\pm1.74$ & $8.224$         & $0.841\pm0.019$ & $5338\pm59$\tablefootmark{b} & $3.877\pm0.084$ & $0.25\pm0.03$  & $1.207\pm0.108$ & $2.029\pm0.286$ & $6.072\pm1.835$  & -5.03  &   $\sim$47 \\[3pt]
	HD 48265  & $11.48\pm0.72$ & $8.051$         & $0.722\pm0.013$ & $5733\pm55$\tablefootmark{d} & $3.970\pm0.048$ & $0.30\pm0.04$  & $1.312\pm0.064$ & $1.901\pm0.126$ & $4.201\pm0.625$  & -5.21  &   $\sim$45 \\[3pt]
	HD 50499  & $21.02\pm0.68$ & $7.207\pm0.004$ & $0.628\pm0.014$ & $6102\pm54$\tablefootmark{b} & $4.247\pm0.033$ & $0.26\pm0.04$  & $1.253\pm0.020$ & $1.351\pm0.054$ & $2.391\pm0.633$  & -5.03  &   $\sim$25 \\[3pt]
	HD 66428  & $18.21\pm1.07$ & $8.246\pm0.013$ & $0.717\pm0.006$ & $5773\pm23$\tablefootmark{b} & $4.360\pm0.050$ & $0.23\pm0.05$  & $1.083\pm0.022$ & $1.102\pm0.062$ & $3.515\pm2.040$  & -5.07  &   $\sim$38 \\[3pt]
	HD 70642  & $34.84\pm0.60$ & $7.169$         & $0.733\pm0.005$ & $5732\pm23$\tablefootmark{b} & $4.458\pm0.017$ & $0.22\pm0.02$  & $1.078\pm0.015$ & $0.984\pm0.022$ & $0.919\pm0.733$  & -4.90  &   $\sim$32 \\[3pt]
	HD 73267  & $18.77\pm1.00$ & $8.889$         & $0.827\pm0.003$ & $5387\pm10$\tablefootmark{d} & $4.447\pm0.035$ & $0.07\pm0.04$  & $0.897\pm0.019$ & $0.909\pm0.033$ & $8.140\pm3.505$  & -4.97  &   $\sim$43 \\[3pt]
	HD 114729 & $27.69\pm0.54$ & $6.687\pm0.005$ & $0.688\pm0.003$ & $5844\pm12$\tablefootmark{c} & $4.046\pm0.016$ & $-0.33\pm0.03$ & $0.936\pm0.013$ & $1.473\pm0.037$ & $10.116\pm0.344$ & -5.07  &   $\sim$34 \\[3pt]
	HD 117207 & $30.37\pm0.92$ & $7.240$         & $0.727\pm0.014$ & $5732\pm53$\tablefootmark{b} & $4.371\pm0.039$ & $0.19\pm0.03$  & $1.053\pm0.028$ & $1.074\pm0.041$ & $4.192\pm2.274$  & -5.06  &   $\sim$39 \\[3pt]
	HD 126525 & $26.24\pm0.82$ & $7.847\pm0.004$ & $0.747\pm0.003$ & $5638\pm13$\tablefootmark{c} & $4.382\pm0.028$ & $-0.10\pm0.01$ & $0.897\pm0.010$ & $0.979\pm0.030$ & $9.670\pm1.333$  & -5.03  &   $\sim$40 \\[3pt]
	HD 143361 & $16.80\pm1.26$ & $9.200$         & $0.792\pm0.004$ & $5507\pm10$\tablefootmark{d} & $4.472\pm0.043$ & $0.14\pm0.06$  & $0.968\pm0.027$ & $0.916\pm0.041$ & $2.942\pm2.749$  & -5.00  &   $\sim$42 \\[3pt]
	HD 152079 & $11.90\pm1.40$ & $9.182$         & $0.687\pm0.014$ & $5907\pm52$                  & $4.365\pm0.054$ & $0.29\pm0.07$  & $1.147\pm0.030$ & $1.128\pm0.074$ & $1.622\pm1.369$  & -4.99  &   $\sim$31 \\[3pt]
	HD 187085 & $22.02\pm1.12$ & $7.225$         & $0.622\pm0.007$ & $6117\pm27$\tablefootmark{b} & $4.279\pm0.041$ & $0.12\pm0.04$  & $1.189\pm0.023$ & $1.270\pm0.066$ & $2.747\pm0.838$  & -4.93  &   $\sim$21 \\[3pt]
	HD 190647 & $18.70\pm1.10$ & $7.779$         & $0.744\pm0.017$ & $5656\pm60$\tablefootmark{b} & $4.162\pm0.054$ & $0.23\pm0.02$  & $1.069\pm0.027$ & $1.376\pm0.089$ & $7.957\pm1.096$  & -5.09  &   $\sim$42 \\[3pt]
	HD 216437 & $37.58\pm0.56$ & $6.057\pm0.001$ & $0.677\pm0.009$ & $5909\pm31$\tablefootmark{b} & $4.188\pm0.026$ & $0.20\pm0.10$  & $1.165\pm0.046$ & $1.394\pm0.032$ & $4.750\pm1.059$  & -5.01  &   $\sim$30 \\[3pt]
	HD 220689 & $21.96\pm0.92$ & $7.795\pm0.062$ & $0.671\pm0.007$ & $5921\pm26$                  & $4.360\pm0.045$ & $-0.07\pm0.10$ & $1.016\pm0.048$ & $1.068\pm0.047$ & $4.586\pm2.487$  & -4.99  &   $\sim$29 \\[3pt]
	\hline
      \end{tabular}
      \tablefoot{All stellar data retrieved from Vizier catalogue, except:
		  \tablefoottext{a}{calculated from PARAM 1.3 \citep[see][]{dasilva2006}}
		  \tablefoottext{b}{retrieved from \citet{bonfanti2015}}
		  \tablefoottext{c}{retrieved from \citet{sousa2008}}
		  \tablefoottext{d}{retrieved from \citet{stassun2017}}
		  \tablefoottext{e}{calculated following \citet{mamajek2008}.}
		 }
    \end{sidewaystable*}
  
    \begin{figure}
      \includegraphics[width=\linewidth]{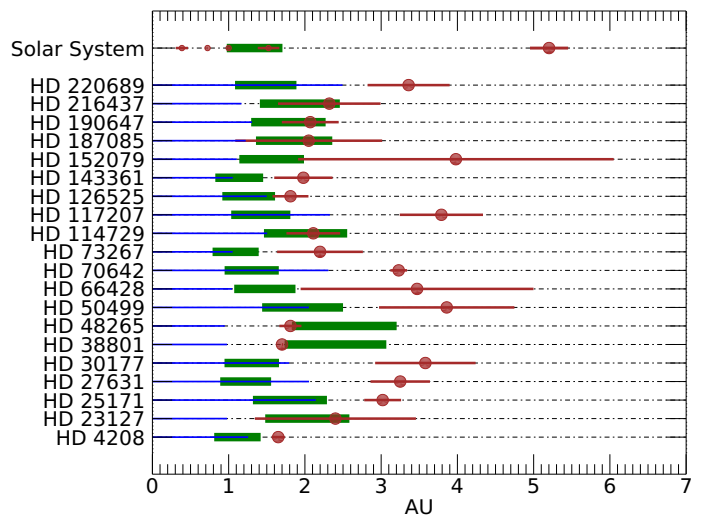}
      \caption{Overview of the sample systems and a comparison with the inner Solar System's architecture. The sample's known giant 
		planets are shown as brown circles, a thin brown line from periastron to apoastron showing their orbit's span. Each 
		system's habitable zone, computed using the model detailed in \citet{kopparapu2013}, is shown as a thick green band, 
		while the thin blue line indicates each system's region of dynamical stability for additional inner planets as 
		computed through Hill's criterion detailed in Sect. \ref{sec:analysis}.}
      \label{fig:sample}
    \end{figure}
    
    The 20 southern-emisphere stars selected for this work are all bright ($V<9.2$), inactive ($\log R^{\prime}_{HK}<-4.8$), not 
    significantly evolved ($\log g>4$) and are not members of visually close binaries, a set of criteria ensuring the overall sample is
    well suited to the search for low-mass inner planets.
    \par All systems feature a single giant planet orbiting the host star at low to moderate eccentricities ($e<0.5$), with 
    periastron larger than 1 AU and discovered via the radial velocity method. Additional details on each system's stellar parameters 
    can be found in Table \ref{table:samplestars}, while a global overview of the planets positions relative 
    to stars and habitable zones is given in Fig. \ref{fig:sample}; all planetary data were retrieved from the most recent published 
    paper discussing each planet's orbital solution and listed in the first column of Tables \ref{table:hd4208} to \ref{table:hd220689}. 
    Regarding stellar 
    parameters, each star's parallax, visual magnitude, metallicity and $\log R^{\prime}_{HK}$ were retrieved from the Vizier online 
    catalogue; effective 
    temperatures were instead retrieved by joining the estimates reported in \citet{bonfanti2015}, \citet{sousa2008} and 
    \citet{stassun2017}. These stellar parameters were then used as input in the online tool PARAM in order to calculate each star's 
    color index, surface gravity, mass, radius and age as detailed in \citet{dasilva2006}. Values of stellar rotational period were 
    instead computed using the empirical relations between $(B-V)$, $\log R^{\prime}_{HK}$ and $P_{rot}$ from \citet{noyes1984} and 
    \citet{mamajek2008} and here reported  with a 1-day precision for the analytical nature of the relations used.
    \par At a general overview of the systems characteristics, it can be noted that all selected stars are reasonably similar to our 
    Sun; two exceptions worthy of note are represented by HD 38801 and HD 48265 whose mass and size (respectively 
    1.36 M$_\odot$, 2.53 R$_\odot$ and 1.28 M$_\odot$, 2.05 R$_\odot$) suggest that they are in a later stage of their evolution, 
    probably on the sub-giant branch.
    \par Moving on to the planets in the chosen sample, the minimum mass range spans from 0.224 M$_J$ to 10.7 M$_J$, the 
    most massive planet being HD 38801 b, which also features peculiar zero values of 
    eccentricity and periastron longitude as fixed in the discovery paper \citep{harakawa2010} to improve its Keplerian fit quality; 
    such a low value of eccentricity at such intermediate distances from the host star ($a=1.7$ AU), it is noted, cannot yet be 
    explained by tidal circularization and therefore represents an interesting conundrum. All the planets in the sample are 
    characterized by long-period orbits, spanning from 1.89 to 7.94 years, and low to moderate eccentricities ($e<0.52$).
    \par It is also important to note that although each selected system was known to host a single, giant planet at the beginning of 
    our observational program, an additional outer giant planet has been recently discovered in the system HD 30177, as detailed in 
    \citet{wittenmeyer2017}. Still, we have nonetheless decided to show rather than discard the results of our observations and analysis 
    on this system, both since the existence of HD 30177 c was not known at the time of observation and because its very long period 
    (P=$31.82$ yr) causes it not to have a significant dynamical influence on the short-period regions of interest to our study.
    \par We further stress the fact 
    that we verified through extensive simulations using literature and archival RV data that current data on the selected sample do 
    not provide the sensitivity needed to succesfully detect Neptune-mass planets with $P<50d$, due to large RV errors ($\geq$5 m/s) 
    in the discovery data, poor sampling at short periods, or both.

  \section{HARPS observations}	\label{sec:observations}
    As discussed in Sect. \ref{sec:introduction}, the present understanding of multiple planetary systems architecture still shows 
    severe gaps, especially regarding the occurence of architectures mimicking that of our Solar System. To try and help filling this 
    gaps we have conducted an intense monitoring on selected stars hosting a single long-period giant planet using the 
    high-precision HARPS spectrograph mounted at the ESO La Silla 3.6m telescope. In order to successfully detect any additional 
    low-mass planets in the selected systems and to further refine the outer giant planet orbit, the selected stars were intensely 
    observed over a time period spanning from April 2014 to March 2015 (ESO programmes 093.C-0919(A) and 094.C-0901(A)); during this 
    period each star was observed for 
    about a semester, and to optimize time sampling the observing nights were split in blocks of two or three nights at the beginning, 
    middle and end of the observing period. Based on the allocated observation time, a total of 514 measurements where obtained during 
    this period, providing about 27 datapoints on average for each of the 20 target stars, a number of observations that, we note, is 
    significantly lower than the $\sim$40 measurements per star that we originally foresaw in the program proposal.
    \par Depending on magnitude, the 
    target stars were measured with 10 or 15 minutes total exposure, in order to achieve sufficient signal-to-noise ratio to reach the 
    required <1 ms$^{-1}$ precision for all program stars using HARPS in its simultaneous Th-Ar observing mode. The aforementioned 
    integration time is also needed in order to average the oscillation and granulation modes of solar-type stars.
    
  \section{Systems analysis}	\label{sec:analysis}
    
    \subsection{Fitting methodology}	\label{subsec:methodology}
      A complete list of the newly acquired high-precision HARPS data for each system in the selected sample is collected in 
      Table \ref{table:data} at the end of this paper. To use this data for the refinement of the known planets orbital 
      elements we subtracted from the raw timeseries the stellar proper radial motion computed as the mean of the radial velocity 
      dataset. Our datasets were then joined with all literature data available on each system, producing therefore a complete series 
      of radial velocity data comprising discovery datasets, follow-up timeseries and the HARPS data acquired during our survey; when 
      available previous HARPS observations publicly released on the ESO archive were also used. Any HARPS datapoint characterized 
      by a signal-to-noise ratio lower than 30 was removed; details on the excluded datapoints are to be found in each system's 
      paragraph in Subsect. \ref{subsec:casebycase}. Unless otherwise noted, we fit for no offset between the various HARPS datasets, 
      that were therefore treated as a single timeseries.
      The main purpose of joining various datasets from different observations and instruments is to use the higher precision of 
      HARPS data to further constrain and refine the known planets orbital parameters by refitting the complete timeseries composed of 
      both historic and new data.
      \par The fits have been obtained using the \texttt{EXOFAST} suite of routines written for IDL \citep{eastman2013}, 
      characterizing the orbital parameters and their uncertainties with a differential evolution Markhov Chain Monte Carlo method, 
      the fitting parameters being time of periastron $T_{peri}$, orbital period $P$, $\sqrt{e}\cos{\omega}$, $\sqrt{e}\sin{\omega}$, 
      semiamplitude $K$ and systematic velocity $\gamma_i$ and jitter $j_i$ for each instrument in the joined timeseries. 
      The \texttt{EXOFAST} routines assumes the stellar jitter to be uncorrelated Gaussian nois that is added in quadrature 
      to the radial velocity measurement error. The known 
      orbital parameters found in the literature and listed in 
      Tables \ref{table:hd4208} to \ref{table:hd220689} were used as starting guesses for the MCMC chains to help achieve a better 
      convergence.
      \par After subtracting the best-fit Keplerian curve produced by \texttt{EXOFAST}, the residual datapoints were used to produce a 
      generalized Lomb-Scargle periodogram using the IDL routine \texttt{GLS} \citep{zechmeister2009} to investigate the presence of 
      additional inner planetary signals, selecting as signals of potential interest for further investigation only those power peaks 
      characterized by a false alarm probabilty as computed via bootstrap method less than or equal to 1\%. 
      \par As a further note on the plausibility of the presence of additional inner planets, after obtaining the refined fits on each 
      system we searched for the maximum orbital period allowing dynamical stability for a single additional planet on a circular inner 
      orbit under the gravitational influence of the known outer planet, using the equality on Hill's criterion 
      \citep[see e.g.][]{murray1999}:
	$$ a_{1}(1-e_{1})-a_{2} \geq 2\sqrt{3}R_H $$
      being $a_{1}$ and $e_{1}$ the major semiaxis and eccentricity of the known giant planet, $a_{2}$ the major semiaxis of the 
      test planet's orbit and $R_H$ the known planet's Hill radius defined as:
	$$ R_H=a_{1}(1-e_{1})\sqrt[3]{\frac{m_{1}}{3 M_*}} $$
      We consider therefore the maximum approach distance between the known outer giant planet and the inner test planet as the  
      limit stability. We varied the test planet's mass in the range 0.1-2 M$_J$ and, since the maximum stable period for the test 
      planet underwent a variation amount to less than one day within this mass range, we provide on each system's paragraph in 
      Subsect. \ref{subsec:casebycase} a mean value for this maximum stable period and show each system's stability region as a thin 
      blue line in Fig. \ref{fig:sample}.
      \par Details on overall results 
      and on case-by-case fitting process and results are contained in the following subsections.
    
    \subsection{Results overview}	\label{subsec:overview}
      The overall results of the new fits obtained joining ours HARPS high-precision data with the available literature time series 
      can be roughly divided in three main categories: fits that confirm the literature without significant changes to the orbital 
      solution, fits than instead significantly change some or all of the published orbital elements and fits that suggest the 
      existence of additional planetary companions in the system. A comparison between the literature results and the parameters 
      obtained from our MCMC chains is shown in each system's Tables \ref{table:hd4208} to \ref{table:hd220689}, along with 
      the number of measurement used for each orbital solution and the mean radial velocity error for each dataset. It is interesting 
      to note that the values for stellar jitter derived from our orbital solutions are often different between the used dataset, 
      the lowest being usually the one found for HARPS datasets; this may suggest that the jitter is likely dominated by instrumental 
      effects for instruments with lower precision than HARPS.
      \par Out of the 20 planetary systems of our sample, 7 did not significantly benefit from the addition of our new, 
      higher-precision HARPS data, resulting in new orbital solution compatible with the ones found in the literature (HD 23127, 
      HD 25171, HD 27631, HD 114729, HD 117207, HD 187085, HD 220689). We however note that for the most part our orbital 
      solutions feature higher precision of the orbital parameters.
      \par Instead, the addition of the HARPS data obtained from our observation brought significant updates in 12 systems of the 
      sample (HD 4208, HD 30177, HD 38801, HD 48265, HD 66428, HD 70642, HD 73267, HD 126525, HD 143361, hd 152079, HD 190647, 
      HD 216437). For the most part these updates consist in values of eccentricity and period incompatible with those found in the 
      literature best-fits solutions; we also want to explicitly stress that our data produced very different orbital parameters 
      for planet HD 30177 c, the addition of orbital parameters missing in the literature fit for HD 126525 and the detection of 
      a linear trend not previously published for HD 73267.
      \par Finally, we report the characterization of outer giant planet HD 50499 c, the existence of which could be previously 
      inferred in the quadratic trend of Keck data but was never fitted and published before.
      
    \subsection{Case-by-case results}	\label{subsec:casebycase}
	\paragraph{HD 4208}
	
	    \begin{table} 
	      \small
		    \caption{Fit results comparison for system HD 4208}\label{table:hd4208}
		    \centering
		    \begin{tabular}{l c c c}
		      \hline\hline
			\multicolumn{4}{c}{HD 4208}\\
		      \hline
						& \multicolumn{2}{c}{\cite{butler2006}}	& This work\\
			Parameter		& \multicolumn{2}{c}{Planet b}		& Planet b\\
		      \hline
			K (ms$^{-1}$)		& \multicolumn{2}{c}{$19.06\pm0.73$}	& $19.03_{-0.79}^{+0.85}$ \\[3pt]
			P (days)		& \multicolumn{2}{c}{$828.0\pm8.1$}	& $832.97_{-1.89}^{+2.15}$ \\[3pt]
			$\sqrt{e}\cos{\omega}$	& \multicolumn{2}{c}{-} 		& $-0.101_{-0.121}^{+0.148}$ \\[3pt]
			$\sqrt{e}\sin{\omega}$	& \multicolumn{2}{c}{-} 		& $-0.092_{-0.123}^{+0.147}$ \\[5pt]
			e			& \multicolumn{2}{c}{$0.052\pm0.040$} 	& $0.042_{-0.029}^{+0.039}$ \\[3pt]
			$\omega$ (deg)		& \multicolumn{2}{c}{$345$} 		& $217.634_{-67.497}^{+57.556}$ \\[3pt]
			M$\sin{i}$ (M$_J$)	& \multicolumn{2}{c}{$0.804\pm0.073$} 	& $0.810_{-0.015}^{+0.014}$ \\[3pt]
			a (AU)			& \multicolumn{2}{c}{$1.650\pm0.096$} 	& $1.662_{-0.015}^{+0.015}$ \\[3pt]
			T$_{peri}$ (d)		& \multicolumn{2}{c}{$2451040\pm120$} 	& $2456505.5_{-491.3}^{+127.3}$ \\[3pt]
		      \hline
			n$_{obs}$		& \multicolumn{2}{c}{52}		& 82\\[3pt]
		      \hline
						& mean RV error	& $\gamma$		& jitter	\\
						& (ms$^{-1}$)	& (ms$^{-1}$)		& (ms$^{-1}$)\\[3pt]
		      \hline
			Keck1		 	& $1.52$	&   $0.12_{-0.58}^{+0.58}$   	& $3.26_{-0.45}^{+0.55}$ \\[3pt]
			Keck2			& $1.44$	&   $-12.39_{-1.99}^{+1.99}$ 	& $6.36_{-1.30}^{+1.86}$ \\[3pt]
			HARPS			& $0.48$	&   $-17.31_{-0.86}^{+0.88}$ 	& $1.33_{-0.19}^{+0.23}$ \\[3pt]
		      \hline
		    \end{tabular}
	    \end{table}
	    \begin{figure}
	      \includegraphics[width=\linewidth]{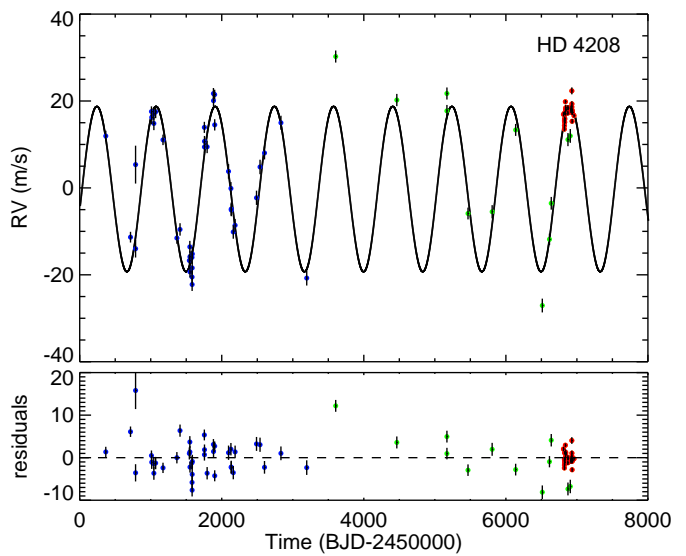}
	      \caption{Analysis results for system HD 4208. In the top panel, our best-fit solution is superimposed as a black curve on the 
			literature datapoints from two Keck surveys (shown in blue and green) and from our HARPS observations (shown in red). 
			The bottom panel shows the residual radial velocities obtained after subtracting pur best-fit solution for HD 4208 b 
			from the original datapoints.}
	      \label{fig:hd4208-fit}
	    \end{figure}
	    
	  This G5V star's only planetary companion was first detected by \citet{vogt2002} via 35 Keck 
	  observations spanning about 5 years. Later observations and improvements in data reduction pipelines reported in 
	  \citet{butler2006} have refined the planet's orbital parameters; additional and further refined Keck data have recently been 
	  published in \citet{butler2017}, although no further orbital refinement has yet been anounced.
	  \par Using all these literature results as historical data next to the 30 high-precision datapoints obtained during our HARPS 
	  survey we obtained a new orbital fit (see Table \ref{table:hd4208} and Fig. \ref{fig:hd4208-fit}) with Keplerian semiamplitude 
	  K=$19.03_{-0.79}^{+0.85}$ms$^{-1}$ and period P=$832.97_{-1.89}^{+2.15}$d, from which we estimate the minimum mass of HD 
	  4208 b to be M$\sin{i}$=$0.810_{-0.015}^{+0.014}$M$_J$, its eccentricity as e=$0.042_{-0.029}^{+0.039}$ and a longitude of 
	  periastron of $\omega$=$217.634_{-67.497}^{+57.556}$deg.
	  \par Our results are generally compatible with Butler's, although better constrained; an exception is represented by a very 
	  different value of $\omega$. No power peak with FAP$\leq$0.01 was found in the residual data periodogram. Using the Hill 
	  criterion we also find the maximum dynamically stable period for an inner planet to be 547.5 days.
	  
	\paragraph{HD 23127}
	
	  \begin{table} 
	    \small
		  \caption{Fit results comparison for system HD 23127}\label{table:hd23127}
		  \centering
		  \begin{tabular}{l c c c}
		    \hline\hline
		      \multicolumn{4}{c}{HD 23127}\\
		    \hline
						& \multicolumn{2}{c}{\cite{otoole2007}}	& This work\\
		      Parameter			& \multicolumn{2}{c}{Planet b}		& Planet b\\
		    \hline
		      K (ms$^{-1}$)		& \multicolumn{2}{c}{$27.5\pm1$} 	& $27.72_{-2.63}^{+2.83}$ \\[3pt]
		      P (days)			& \multicolumn{2}{c}{$1214\pm9$} 	& $1211.17_{-8.91}^{+11.11}$ \\[3pt]
		      $\sqrt{e}\cos{\omega}$	& \multicolumn{2}{c}{-}			& $-0.613_{-0.065}^{+0.088}$ \\[3pt]
		      $\sqrt{e}\sin{\omega}$	& \multicolumn{2}{c}{-}			& $-0.019_{-0.185}^{+0.176}$ \\[3pt]
		      e				& \multicolumn{2}{c}{$0.44\pm0.07$} 	& $0.406_{-0.09}^{+0.083}$ \\[3pt]
		      $\omega$ (deg)		& \multicolumn{2}{c}{$190\pm6$} 	& $181.829_{-16.507}^{+17.296}$ \\[3pt]
		      M$\sin{i}$ (M$_J$)	& \multicolumn{2}{c}{$1.5\pm0.2$} 	& $1.527_{-0.038}^{+0.037}$ \\[3pt]
		      a (AU)			& \multicolumn{2}{c}{$2.4\pm0.3$} 	& $2.370_{-0.032}^{+0.032}$ \\[3pt]
		      T$_{peri}$ (days)		& \multicolumn{2}{c}{$2450229\pm19$} 	& $2457266.6_{-43.5}^{+62.1}$ \\[3pt]
		    \hline
			n$_{obs}$		& \multicolumn{2}{c}{34}		&58\\[3pt]
		    \hline
						& mean RV error	& $\gamma$		& jitter			\\
						& (ms$^{-1}$)	& (ms$^{-1}$)		& (ms$^{-1}$)\\[3pt]
		    \hline
		      AAT			& 4.30		& $5.55_{-2.27}^{+2.32}$ & $11.00_{-1.60}^{+2.01}$\\[3pt]
		      HARPS			& 0.64		& $-1.82_{-4.53}^{+4.69}$& $1.88_{-0.29}^{+0.38}$ \\[3pt]
		    \hline
		  \end{tabular}
	  \end{table}
	  \begin{figure}
	    \includegraphics[width=\linewidth]{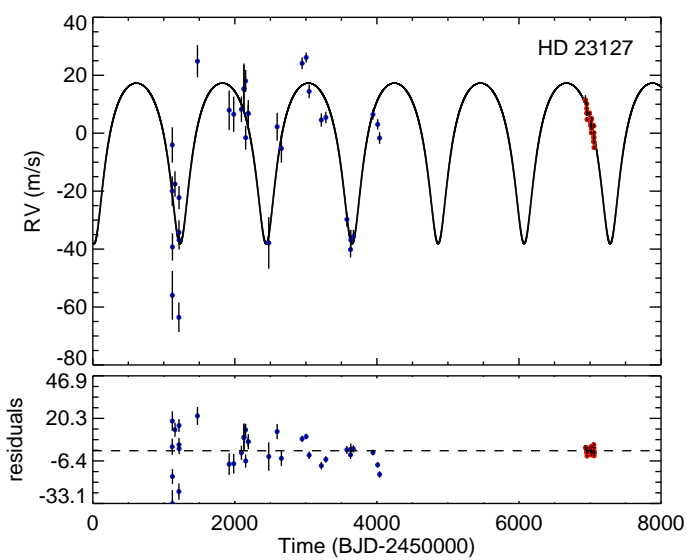}
	    \caption{Same as Fig. \ref{fig:hd4208-fit} but for system HD 23127. The literature AAT data are shown in blue, while our HARPS 
		      survey data are shown in red.}
	    \label{fig:hd23127-fit}
	  \end{figure}
	  
	  A G2V star, its planetary companion has been announced in \citet{otoole2007} based on 34 observations 
	  collected at the Anglo-Australian Telescope (AAT) over 7.7 yr; no further data nor refinement to the planet's orbital 
	  elements have been published since its discovery.
	  \par The HARPS observations have produced 25 new datapoint; we excluded a single datapoint taken at epoch $2456989.5$ 
	  since it showed a radial velocity difference from the observation taken the previous and following nights of 
	  $\sim$30 ms$^{-1}$, in addition to having an outlier value of bisector; being such a difference difficult to justify as a
	  real radial velocity variation the point was flagged as spurious, probably due to some error occurring during its acquisition, 
	  and therefore ignored in the subsequent analysis.
	  \par The orbital fit (see Table \ref{table:hd23127} and Fig. \ref{fig:hd23127-fit}) obtained from joining the 34 AAT 
	  datapoints with the 24 HARPS observations 
	  suggests new orbital elements values of M$\sin{i}$=$1.527_{-0.038}^{+0.037}$M$_J$, P=$1211.17_{-8.91}^{+11.11}$d and 
	  e=$0.406_{-0.09}^{+0.083}$, results compatible to the literature solution. The radial velocity residuals periodogram show no 
	  power peak under the 0.01 level of FAP, and we find the maximum stable period for an additional inner planet to be 322.1 days.
    
	\paragraph{HD 25171}
	
	  \begin{table} 
	    \small
		  \caption{Fit results comparison for system HD 25171}\label{table:hd25171}
		  \centering
		  \begin{tabular}{l c c c}
		    \hline\hline
		      \multicolumn{4}{c}{HD 25171}\\
		    \hline
						& \multicolumn{2}{c}{\cite{moutou2011}}& This work\\
		      Parameter			& \multicolumn{2}{c}{Planet b}		& Planet b\\
		    \hline
		      K (ms$^{-1}$)		& \multicolumn{2}{c}{$15.0\pm3.6$} 	& $14.56_{-0.88}^{+0.84}$ \\[3pt]
		      P (days)			& \multicolumn{2}{c}{$1845\pm167$} 	& $1802.29_{-22.92}^{+24.12}$ \\[3pt]
		      $\sqrt{e}\cos{\omega}$	& \multicolumn{2}{c}{-}			& $-0.006_{-0.169}^{+0.175}$ \\[3pt]
		      $\sqrt{e}\sin{\omega}$	& \multicolumn{2}{c}{-}			& $0.028_{-0.169}^{+0.159}$ \\[3pt]
		      e				& \multicolumn{2}{c}{$0.08\pm0.06$} 	& $0.042_{-0.029}^{+0.046}$ \\[3pt]
		      $\omega$ (deg)		& \multicolumn{2}{c}{$96\pm89$} 	& $159.543_{-104.900}^{+136.626}$ \\[3pt]
		      M$\sin{i}$ (M$_J$)	& \multicolumn{2}{c}{$0.95\pm0.10$} 	& $0.915_{-0.012}^{+0.011}$ \\[3pt]
		      a (AU)			& \multicolumn{2}{c}{$3.02\pm0.16$} 	& $2.971_{-0.032}^{+0.033}$ \\[3pt]
		      T$_{peri}$ (days)		& \multicolumn{2}{c}{$2455301\pm449$} 	& $2457105.7_{-554.43}^{+540.5}$ \\[3pt]
		    \hline
			n$_{obs}$		& \multicolumn{2}{c}{24}		& 46 \\[3pt]
		    \hline
						& mean RV error	& $\gamma$		& jitter			\\
						& (ms$^{-1}$)	& (ms$^{-1}$)		& (ms$^{-1}$)\\[3pt]
		    \hline
		      HARPS			&  1.18		& $-2.10_{-0.58}^{+0.60}$ & $2.40_{-0.34}^{+0.39}$ \\[3pt]
		    \hline
		  \end{tabular}
	  \end{table}
	  \begin{figure}
	    \includegraphics[width=\linewidth]{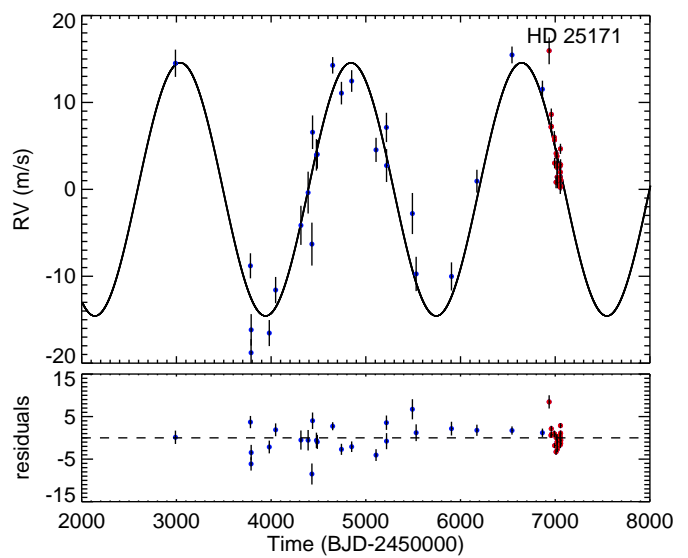}
	    \caption{Same as Fig. \ref{fig:hd4208-fit} but for system HD 25171. The literature HARPS data are shown in blue, while our HARPS 
		      survey data are shown in red.}
	    \label{fig:hd25171-fit}
	  \end{figure}
	
	  This F8V-type star's planet has been discovered and characterized in \citet{moutou2011} via 24 HARPS 
	  observations; 8 additional HARPS datapoints are publicly accessible from the ESO Science Archive. Of these 32 archival 
	  HARPS data we excluded a total of 8 datapoints due to SNR lower than 30, namely the data found at epochs 2452946.79, 
	  2452998.59, 2454732.77, 2454813.77, 2455091.89, 2455105.77, 2455261.53 and 2456751.48.
	  \par In addition to this literature HARPS data we used our own 22 lower-uncertainties datapoints to obtain a 
	  K=$14.56_{-0.88}^{+0.85}$, P=$1802.29_{-22.92}^{+24.12}$d Keplerian fit (see Table \ref{table:hd25171} and 
	  Fig. \ref{fig:hd25171-fit}) from which we provide new estimates of minimum mass of $0.915_{-0.012}^{+0.012}$M$_J$ and 
	  eccentricity of $0.0417_{-0.029}^{+0.046}$, in a generally good agreement with the literature fit. No power peak with a 
	  sufficiently low false alarm probability is present in the residuals periodogram. The maximum stable period for an inner 
	  planet found via the Hill criterion is 1105 days.
    
	\paragraph{HD 27631}
	  
	  \begin{table} 
	    \small
		  \caption{Fit results comparison for system HD 27631}\label{table:hd27631}
		  \centering
		  \begin{tabular}{l c c c}
		    \hline\hline
		      \multicolumn{4}{c}{HD 27631}\\
		    \hline
						& \multicolumn{2}{c}{\cite{marmier2013}}& This work\\
		      Parameter			& \multicolumn{2}{c}{Planet b}		& Planet b\\
		    \hline
		      K (ms$^{-1}$)		& \multicolumn{2}{c}{$23.7\pm1.9$}	& $24.51_{-1.82}^{+1.84}$ \\[3pt]
		      P (days)			& \multicolumn{2}{c}{$2208\pm66$}	& $2198.14_{-50.34}^{+54.11}$ \\[3pt]
		      $\sqrt{e}\cos{\omega}$	& \multicolumn{2}{c}{-}			& $-0.182_{-0.143}^{+0.177}$ \\[3pt]
		      $\sqrt{e}\sin{\omega}$	& \multicolumn{2}{c}{-}			& $0.284_{-0.159}^{+0.113}$ \\[3pt]
		      e				& \multicolumn{2}{c}{$0.12\pm0.060$}	& $0.141_{-0.065}^{+0.062}$ \\[3pt]
		      $\omega$ (deg)		& \multicolumn{2}{c}{$134\pm44$} 	& $122.642_{-30.158}^{+33.335}$ \\[3pt]
		      M$\sin{i}$ (M$_J$)	& \multicolumn{2}{c}{$1.45\pm0.14$} 	& $1.494_{-0.042}^{+0.042}$ \\[3pt]
		      a (AU)			& \multicolumn{2}{c}{$3.25\pm0.07$} 	& $3.242_{-0.068}^{+0.070}$ \\[3pt]
		      T$_{peri}$ (days)		& \multicolumn{2}{c}{$2453867\pm224$} 	& $2456086.0_{-170.8}^{+193.4}$ \\[3pt]
		    \hline
			n$_{obs}$		& \multicolumn{2}{c}{64}		&87\\[3pt]
		    \hline
						& mean RV error	& $\gamma$		& jitter			\\
						& (ms$^{-1}$)	& (ms$^{-1}$)		& (ms$^{-1}$)\\[3pt]
		    \hline
		      CORALIE1			& 5.84		& $8.19_{-3.28}^{+3.24}$& $5.29_{-1.74}^{+1.97}$ \\[3pt]
		      CORALIE2			& 3.76		& $-3.74_{-1.68}^{+1.59}$& $7.42_{-1.05}^{+1.19}$ \\[3pt]
		      HARPS			& 0.53		& $-1.79_{-3.47}^{+3.31}$& $4.24_{-0.61}^{+0.79}$ \\[3pt]
		    \hline
		  \end{tabular}
	  \end{table}
	  \begin{figure}
	    \includegraphics[width=\linewidth]{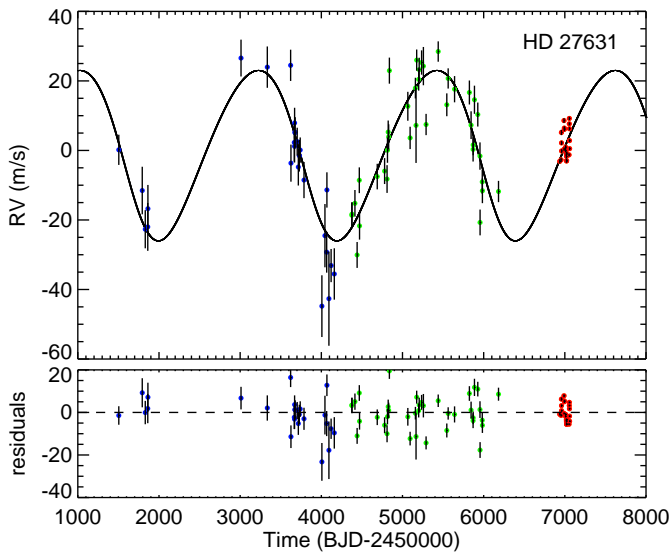}
	    \caption{Same as Fig. \ref{fig:hd4208-fit} but for system HD 27631. The literature CORALIE surveys data are shown in blue and 
		      green, while our HARPS survey data are shown in red.}
	    \label{fig:hd27631-fit}
	  \end{figure}
	  
	  The planet orbiting this G3IV star was first discovered by \citet{marmier2013} using a total of 64 
	  CORALIE measurements, the last 37 of which were obtained after an upgrade of the instrument and are therefore treated as a 
	  separate dataset with its own radial velocity offset.
	  \par During our HARPS observations we have obtained 23 further datapoints; the joined time series was then used to produce 
	  a Keplerian fit having semiamplitude K=$24.51_{-1.82}^{+1.84}$ms$^{-1}$ and period P=$2198.14_{-50.34}^{+54.11}$d, providing 
	  a new estimate of M$\sim{i}$=$1.494_{-0.042}^{+0.042}$M$_J$ and eccentricity e=$0.141_{-0.065}^{+0.062}$, an overall solution 
	  compatible yet better constrained than the literature fit (see Table \ref{table:hd27631} and Fig. \ref{fig:hd27631-fit}). 
	  We also find the maximum stable period for an additional inner planet to be 1107 days.
	  \par The generalized Lomb-Scargle periodogram produced from the residual data features no major peak with FAP$\le$0.01.
	
	\paragraph{HD 30177}
	
	  \begin{table*} 
	    \small
		  \caption{Fit results comparison for system HD 30177}\label{table:hd30177}
		  \centering
		  \begin{tabular}{l c c c c}
		    \hline\hline
		      \multicolumn{5}{c}{HD 30177}\\
		    \hline
					        & \cite{wittenmeyer2017}& This work 		   & \cite{wittenmeyer2017}& This work\\
		      Parameter			& Planet b		& Planet b		   & Planet c		&Planet c\\
		    \hline
		      K (ms$^{-1}$)		& $126.3\pm1.5$		& $125.98_{-1.27}^{+1.26}$ & $70.8\pm29.5$ 	& $\geq16.42$ \\[3pt]
		      P (days)			& $2524.4\pm9.8$ 	& $2527.83_{-4.69}^{+4.71}$& $11613\pm1837$ 	& $\geq6163.40$ \\[3pt]
		      $\sqrt{e}\cos{\omega}$	& -			& $0.340_{-0.014}^{+0.014}$& -			& - \\[3pt]
		      $\sqrt{e}\sin{\omega}$	& -			& $0.216_{-0.016}^{+0.015}$& -			& - \\[3pt]
		      e				& $0.184\pm0.012$	& $0.162_{-0.010}^{+0.010}$& $0.22\pm0.14$	& - \\[3pt]
		      $\omega$ (deg)		& $31\pm3$ 		&$32.422_{-2.563}^{+2.525}$& $19\pm30$ 		& - \\[3pt]
		      M$\sin{i}$ (M$_J$)	& $8.08\pm0.10$ 	& $8.622_{-0.126}^{+0.125}$& $7.6\pm3.1$ 	& $\geq1.533$ \\[3pt]
		      a (AU)			& $3.58\pm0.01$		& $3.704_{-0.027}^{+0.027}$& $9.89\pm1.04$	& - \\[3pt]
		      T$_{peri}$ (days)		& $2451434\pm29$	& $2456502.4_{-18.4}^{+17.9}$& $2448973\pm1211$	& - \\[3pt]
		    \hline
			n$_{obs}$		& 63			& 85			& &\\[3pt]
		    \hline
						& mean RV error	& $\gamma$		& jitter	&		\\
						& (ms$^{-1}$)	& (ms$^{-1}$)		& (ms$^{-1}$)	&		\\[3pt]
		    \hline
		      AAT			& 3.84		& $22.16_{-1.73}^{+1.69}$  & $7.43_{-1.17}^{+1.38}$			&  \\[3pt]
		      HARPS			& 0.79		& $-6.66_{-1.18}^{+1.20}$  & $2.19_{-0.34}^{+0.39}$			&  \\[3pt]
		    \hline
		  \end{tabular}
	  \end{table*}
	  
	  The planetary system orbiting this G8V star is a well-studied one and therefore deserves special 
	  attention. The first study to detect a planet around the star \citep{tinney2003} used fifteen AAT observations and 
	  failed to observe the planet's entire orbit, leading to noticeably poor constrained values of e=$0.22\pm0.17$ and 
	  M$\sin{i}$=$7.7\pm1.5$M$_J$. A follow-up study \citep{butler2006} refines these values leading to estimates of 
	  e=$0.193\pm0.025$ and a minimum mass of $10.45\pm0.88$M$_J$. Most recently, \citet{wittenmeyer2017} reports an orbital 
	  fit made from 28 additional AAT data and 20 HARPS-TERRA measurements, leading to the latest values of 
	  M$\sin{i}$=$8.08\pm0.10$M$_J$, P=$2524.4\pm9.8$d, e=$0.184\pm0.012$; of particular interest is the fact that the 
	  authors best-fit solution feature an additional, outer planetary companion HD 30177 c, estimating a minimum mass of 
	  $7.60\pm3.10$M$_J$, a period of $11613\pm1837$ days and an eccentricity value of $0.22\pm0.14$.
	  \par As already noted and discussed in Sect. \ref{sec:sample}, the existence of planet c was not known during our targets 
	  selection and subsequent HARPS observations, hence the inclusion of this system in an otherwise single-planet sample. We 
	  however decided to still include HD 30177 in this paper, since we consider the published outer companion's long period of 
	  31.82 yr not to have a significant dynamical influence in the stability of the short-period (a<1 AU) region of interest in 
	  searching for inner, low-mass companions. We therefore report here the result of the fit obtained joining the literature data 
	  to the additional 26 HARPS measurements collected in our observations, noting that we excluded from the following 
	  analysis four datapoints taken at epochs 2454384.87, 2455563.54, 2455564.57 and 2456631.80 due to low signal-to-noise ratio.
	  \par It can be noted that the dataset used in \citet{wittenmeyer2017} fails to cover a significant portion of the 
	  orbit of 
	  HD 30177 c, covering a total timescale of about only 5840 days out of the 31.7 years proposed orbital period; a fact clearly 
	  reflected in the poorly constrained orbital parameters for this proposed outer planet. Moreover, all the MCMC chains we 
	  launched in searching for a two-planet solution failed to achieve a 
	  satisfactory convergence and parameters resolutions, the acceptance rate never rising above 0.4\% and producing 
	  exceptionally poorly constrained estimates for the outer companion's orbital parameters, such as a period of 
	  $11084.29_{-1059.33}^{+1235.54}$ days and a semiamplitude of $32.71_{-26.48}^{+83.55}$ ms$^{-1}$.
	  \par We interpret the poor parameter 
	  constrains in \citet{wittenmeyer2017} and our repeated failure to achieve a satisfactory convergence as evidence that the 
	  most recent literature two-planet fit does not constitute a correct interpretation of the available data, while an orbital 
	  solution featuring the outer planet as a quadratic trend added to the Keplerian signal of the inner planet may be a better 
	  fit to the joined dataset, allowing a further refinement of the orbital parameters for HD 30177 b but only lower limits on 
	  the outer planet's semiamplitude, mass and period. 
	  \par
\begin{figure}
	    \includegraphics[width=\linewidth]{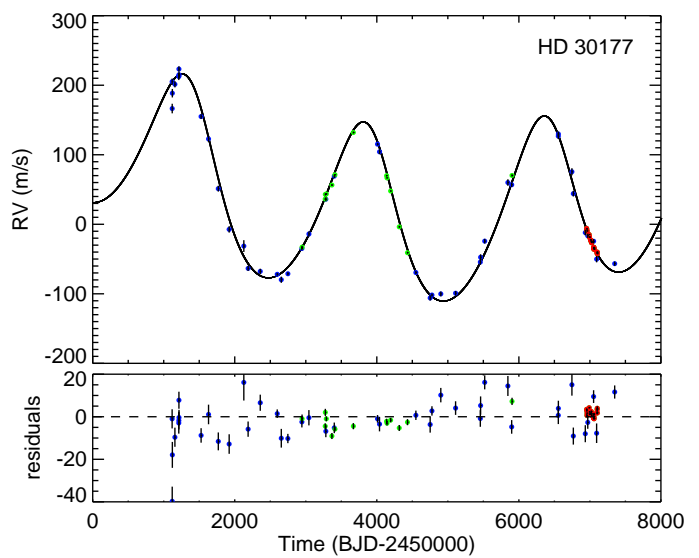}
	    \caption{Same as Fig. \ref{fig:hd4208-fit} but for system HD 30177. The literature AAT data are shown in blue and the 
		      literature HARPS data are shown in green; our HARPS survey data are instead shown in red.}
	    \label{fig:hd30177-fit}
	  \end{figure}

 Following the example set by \citet{kipping2011} we therefore fitted the radial velocity data for a single Keplerian 
	  orbit plus a quadratic term, so that the overall RV model is expressed as:
	    $$ RV=V_\gamma-K\sin{\frac{2\pi(t-t_{peri})}{P}}+k_2(t-t_{pivot})+\frac{1}{2}k_3(t-t_{pivot})^2 $$
	  being $t_{pivot}$ the mean time stamp and $k_2$, $k_3$ respectively the linear and quadratic coefficients. From this model 
	  we can infer some estimate on minimum mass and period for the outer planet using the relationships:
	    $$ \frac{-k_2+k_3 t_{pivot}}{k_3}=t_{peri}+\frac{P}{4} \quad;\quad \frac{k_3}{4\pi^2}=\frac{K}{P^2} $$
	  Our best-fit curve (see Table \ref{table:hd30177} and Fig. \ref{fig:hd30177-fit}) returns for planet b values for 
	  semiamplitude 
	  $125.98_{-1.27}^{+1.26}$ms$^{-1}$ ,minimum mass of $8.622_{-0.126}^{+0.125}$ M$_J$, period  $2527.83_{-4.69}^{+4.71}$ d 
	  and eccentricity $0.162_{-0.010}^{+0.010}$; we find quadratic trend coefficients of 
	  $k_2$=$(-6.7\pm0.5)\cdot10^{-3}$ms$^{-1}$d$^{-1}$ and $k_2$=$(1.20\pm0.05)\cdot10^{-5}$ms$^{-1}$d$^{-1}$. Again following 
	  \citet{kipping2011}, to determine the lower limits 
	  for the outer planet's semiamplitude, mass and period we force a circular orbit, selecting as minimum value of P the 
	  one causing a $\delta\chi^2=1$ relative to the quadratic fit, and then using this period value for computing a minimum value 
	  of K and then minimum mass. The limits thus obtained for the outer planet are P$\geq6163.40$d, K$\geq16.42$ ms$^{-1}$ and 
	  M$\sin{i}\geq1.533$ M$_J$; further datapoints and a greater observation coverage are clearly necessary to better characterize 
	  this outer companion.
	  \par The most significant peak apparent in the residual data periodogram is located at 771.0 days and has a false alarm 
	  probability of $0.1$, above our threshold of 1\%. The maximum orbital period allowing dynamical stability for an additional 
	  inner planet obtained from Hill's criterion is 854.5 days.
	    
	\paragraph{HD 38801}
	  
	  This G8IV star is host to a noticeable case of low-eccentricity super-massive planet discovered by 
	  \citet{harakawa2010}, deserving a short literature retrospective. Basing its detection on 21 Subaru and Keck joint 
	  observations, the authors reported the results of a Keplerian fit with semiamplitude $200.0\pm3.9$ms$^{-1}$ and period of 
	  $696.3\pm2.7$ days, obtaining a minimum mass of $10.7\pm0.5$M$_J$ and an extraordinary low eccentricity, set to zero 
	  after its first estimate of 0.04 led to poorly constrained value of $\omega$, also set to zero to improve the fit. 
	  HD 38801 b's low eccentricity, the authors note, is of special interest since it cannot be explained by tidal 
	  interaction with its host star since the latter's radius of 2.5 R$_\odot$ is too small to effectively circularize the 
	  planet's intermediate orbit.
	  \par
\begin{figure}
	    \includegraphics[width=\linewidth]{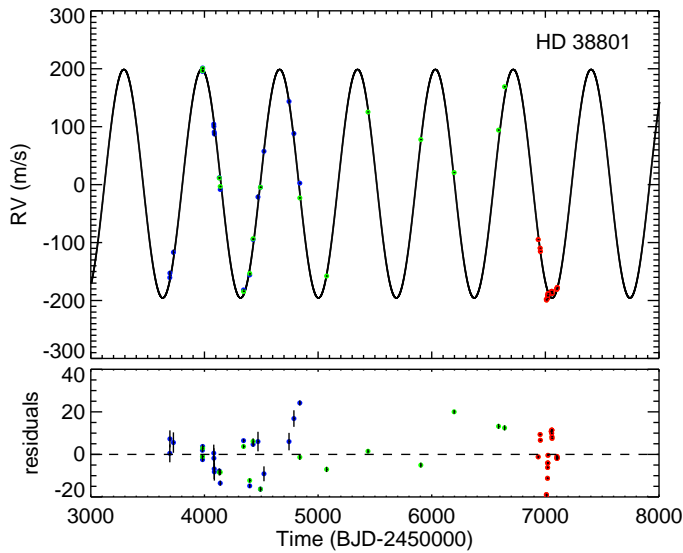}
	    \caption{Same as Fig. \ref{fig:hd4208-fit} but for system HD 38801. The literature Subaru data are shown in blue, literature 
		      Keck data is in green, while our HARPS survey data are shown in red.}
	    \label{fig:hd38801-fit}
	  \end{figure}

 The addition of 16 Keck measurments published in \citet{butler2017} and of our own 21 HARPS measurements allow us to 
	  produce a new best-fit solution (see Table \ref{table:hd38801} and Fig. \ref{fig:hd38801-fit}) with semiamplitude 
	  $197.29_{-3.43}^{+3.52}$ms$^{-1}$ and 
	  period $685.25_{-0.85}^{+0.85}$d, obtaining a minimum mass value of $9.698_{-0.587}^{+0.573}$M$_J$ and 
	  still poorly constrained values for e=$0.017_{-0.011}^{+0.013}$ and $\omega$=$97.115_{-71.900}^{+238.614}$deg that will 
	  surely benefit from further high-precision and high-sampling observations. We find the dynamical stability limit period for 
	  additional inner planets to be 322.1 days.
	  \par
\begin{figure}
      \includegraphics[width=\linewidth]{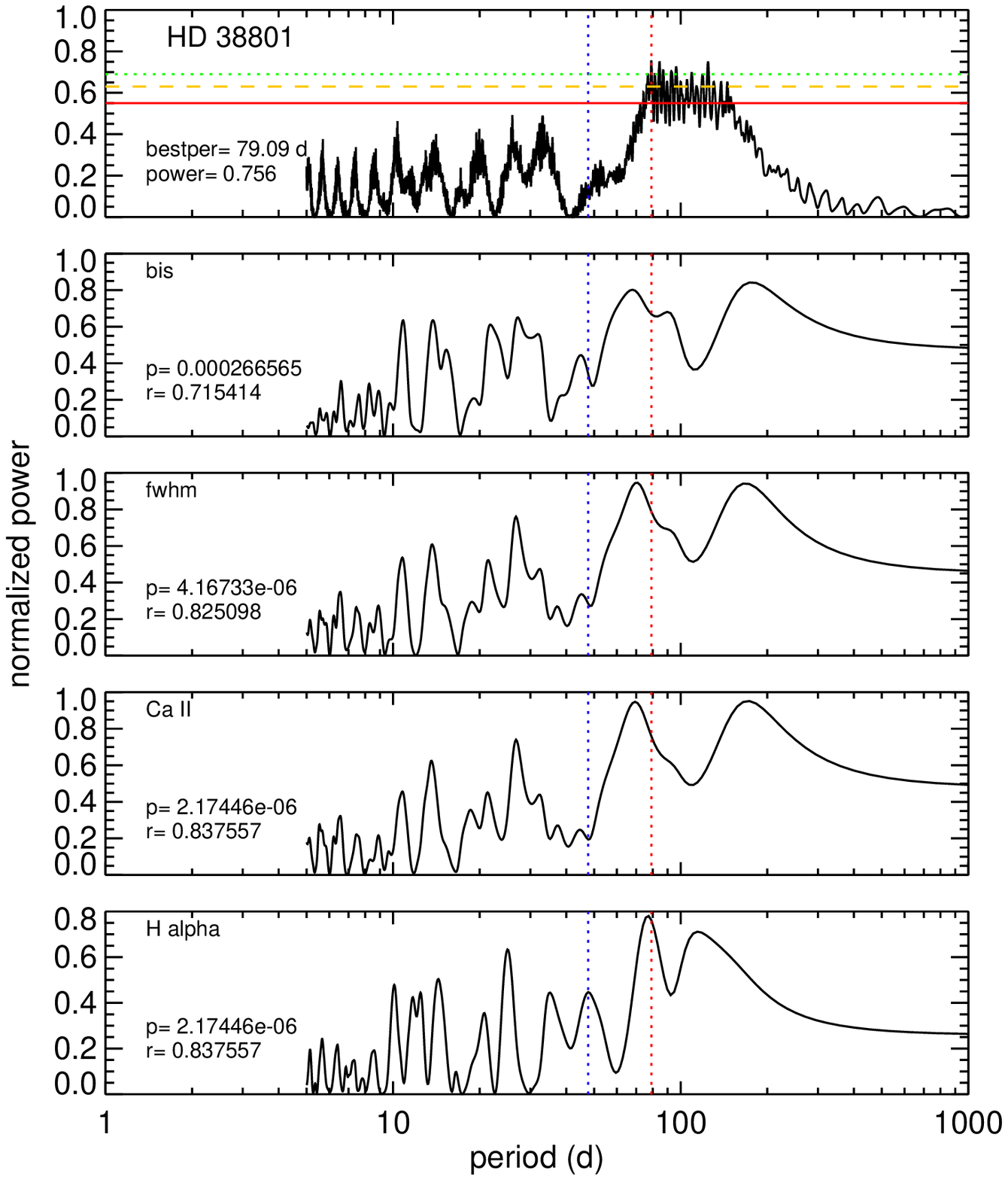}
      \caption{Activity indexes periodograms for star HD 38801. The topmost panel shows the post-fit residual data periodogram, the 
		most significant period peak value and power 
		shown in the lower left corner, the horizontal lines indicating the false alarm probability levels of 10\% 
		(solid red line), 1\% (dashed orange line) and 0.1\% (dotted green line). The following panels show the periodograms 
		for the activity indexes studied in the 
		paper, respectively bisector, FWHM, Ca II and H$\alpha$, the correlation rank $r$ and significance $p$ with the radial 
		velocity residuals shown in each panel's lower left corner. The residual data periodogram's most significant period is 
		highlighted by a vertical red dotted line in all the panels, while the stellar rotation period is shown as a blue 
		dotted vertical line.}
      \label{fig:hd38801-activity}
    \end{figure}

 A power peak characterized by a very low FAP of 0.01\% is present in the residual radial velocity periodogram around 
	  79 days; as shown in Fig. \ref{fig:hd38801-activity} this peak has however a high correlation with similar peaks in the 
	  stellar activity indexes, excluding any planetary origin for this signal.
	  
	  \begin{table} 
	    \small
		  \caption{Fit results comparison for system HD 38801}\label{table:hd38801}
		  \centering
		  \begin{tabular}{l c c c}
		    \hline\hline
		      \multicolumn{4}{c}{HD 38801}\\
		    \hline
						& \multicolumn{2}{c}{\cite{harakawa2010}}& This work\\
		      Parameter			& \multicolumn{2}{c}{Planet b}		& Planet b\\
		    \hline
		      K (ms$^{-1}$)		& \multicolumn{2}{c}{$200.0\pm3.9$}	& $197.29_{-3.43}^{+3.52}$ \\[3pt]
		      P (days)			& \multicolumn{2}{c}{$696.3\pm2.7$} 	& $685.25_{-0.85}^{+0.85}$ \\[3pt]
		      $\sqrt{e}\cos{\omega}$	& \multicolumn{2}{c}{-}			& $0.089_{-0.086}^{+0.058}$ \\[3pt]
		      $\sqrt{e}\sin{\omega}$	& \multicolumn{2}{c}{-}			& $0.021_{-0.093}^{+0.087}$ \\[3pt]
		      e				& \multicolumn{2}{c}{$0.0$} 		& $0.017_{-0.011}^{+0.013}$ \\[3pt]
		      $\omega$ (deg)		& \multicolumn{2}{c}{$0.0$} 		& $97.115_{-71.900}^{+238.614}$ \\[3pt]
		      M$\sin{i}$ (M$_J$)	& \multicolumn{2}{c}{$10.7\pm0.5$} 	& $9.698_{-0.587}^{+0.573}$ \\[3pt]
		      a (AU)			& \multicolumn{2}{c}{$1.7\pm0.037$} 	& $1.623_{-0.049}^{+0.047}$ \\[3pt]
		      T$_{peri}$ (days)		& \multicolumn{2}{c}{$2453966.0\pm2.1$}	& $2456752.9_{-102.1}^{+126.5}$ \\[3pt]
		    \hline
			n$_{obs}$		& \multicolumn{2}{c}{37}		& 58\\[3pt]
		    \hline
						& mean RV error	& $\gamma$		& jitter			\\
						& (ms$^{-1}$)	& (ms$^{-1}$)		& (ms$^{-1}$)\\[3pt]
		    \hline
		      Subaru			& 2.70		& $1.44_{-2.89}^{+2.96}$   & $10.81_{-1.71}^{+2.20}$ \\[3pt]
		      Keck			& 0.50		& $-20.85_{-2.95}^{+2.87}$ & $10.93_{-1.98}^{+2.76}$ \\[3pt]
		      HARPS			& 0.45		& $175.71_{-4.77}^{+4.63}$ & $10.13_{-1.58}^{+2.05}$ \\[3pt]
		    \hline
		  \end{tabular}
	  \end{table}
	
	\paragraph{HD 48265}
	
	  \begin{table} 
	    \small
		  \caption{Fit results comparison for system HD 48265}\label{table:hd48265}
		  \centering
		  \begin{tabular}{l c c c}
		    \hline\hline
		      \multicolumn{4}{c}{HD 48265}\\
		    \hline
						& \multicolumn{2}{c}{\cite{jenkins2017}}	& This work\\
		      Parameter			& \multicolumn{2}{c}{Planet b}		& Planet b\\
		    \hline
		      K (ms$^{-1}$)		& \multicolumn{2}{c}{$27.7\pm1.2$}	& $28.65_{-1.57}^{+1.59}$ \\[3pt]
		      P (days)			& \multicolumn{2}{c}{$780.3\pm4.6$} 	& $778.51_{-5.18}^{+5.38}$ \\[3pt]
		      $\sqrt{e}\cos{\omega}$	& \multicolumn{2}{c}{-}			& $0.298_{-0.123}^{+0.104}$ \\[3pt]
		      $\sqrt{e}\sin{\omega}$	& \multicolumn{2}{c}{-}			& $-0.319_{-0.123}^{+0.224}$ \\[3pt]
		      e				& \multicolumn{2}{c}{$0.080\pm0.050$} 	& $0.211_{-0.096}^{+0.089}$ \\[3pt]
		      $\omega$ (deg)		& \multicolumn{2}{c}{$343.78\pm137.51$} & $308.313_{-23.111}^{+21.261}$ \\[3pt]
		      M$\sin{i}$ (M$_J$)	& \multicolumn{2}{c}{$1.47\pm0.12$} 	& $1.525_{-0.050}^{+0.049}$ \\[3pt]
		      a (AU)			& \multicolumn{2}{c}{$1.81\pm0.07$} 	& $1.814_{-0.031}^{+0.030}$ \\[3pt]
		      T$_{peri}$ (days)		& \multicolumn{2}{c}{$2452892\pm100$} 	& $2456036.7_{-36.5}^{+67.7}$ \\[3pt]
		    \hline
			n$_{obs}$		& \multicolumn{2}{c}{60}		& 20 \\[3pt]
		    \hline
						& mean RV error	& $\gamma$		& jitter			\\
						& (ms$^{-1}$)	& (ms$^{-1}$)		& (ms$^{-1}$)\\[3pt]
		    \hline
		      MIKE			& 3.22		& $-5.81_{-2.08}^{+2.29}$ & $3.38_{-1.53}^{+1.89}$ \\[3pt]
		      HARPS			& 0.54		& $-6.37_{-2.72}^{+2.68}$ & $3.38_{-1.53}^{+1.89}$ \\[3pt]
		      CORALIE			& 9.40		& $-1.23_{-4.49}^{+4.45}$ & $15.98_{-3.21}^{+3.96}$ \\[3pt]
		    \hline
		  \end{tabular}
	  \end{table}
	  \begin{figure}
	    \includegraphics[width=\linewidth]{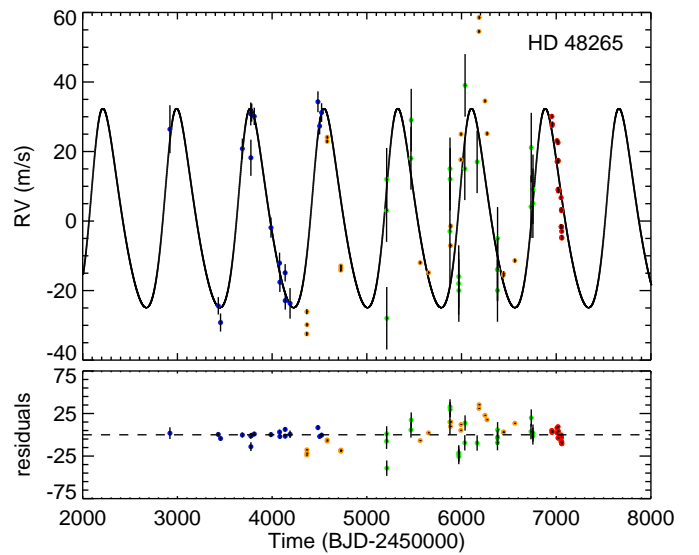}
	    \caption{Same as Fig. \ref{fig:hd4208-fit} but for system HD 48265. The literature MIKE data are shown in blue, the literature 
		      HARPS data in orange and literature CORALIE datapoints is in green, while our HARPS survey data are shown in red.}
	    \label{fig:hd48265-fit}
	  \end{figure}
	  
	  This G5IV star hosts a planet first characterized in \citet{minniti2009} via the analysis of 17 
	  observations obtained over 4.4 years with the MIKE echelle spectrograph, and later refined using new HARPS  and CORALIE data 
	  in \citet{jenkins2009,jenkins2017}.
	  \par Our 20 new HARPS observations, when joined with the literature data, lead to a fit characterized by semiamplitude 
	  K=$28.65_{-1.57}^{+1.59}$ms$^{-1}$ and period P=$778.51_{-5.18}^{+5.38}$d. From this fit (see Table \ref{table:hd48265} and 
	  Fig. \ref{fig:hd48265-fit}) we 
	  obtain orbital elements M$\sin{i}$=$1.525_{-0.050}^{+0.049}$ and e=$0.211_{-0.096}^{+0.089}$. Although most of these fit parameters are reasonably consistent with those found by \citet{jenkins2017}, 
	  we note that our estimate of eccentricity is only marginally compatible with the one obtained in the published works. The 
	  maximum inner period allowing dynamical stability for additional inner planets is found with Hill's criterion to be 294.7 days.
	  \par
\begin{figure}
      \includegraphics[width=\linewidth]{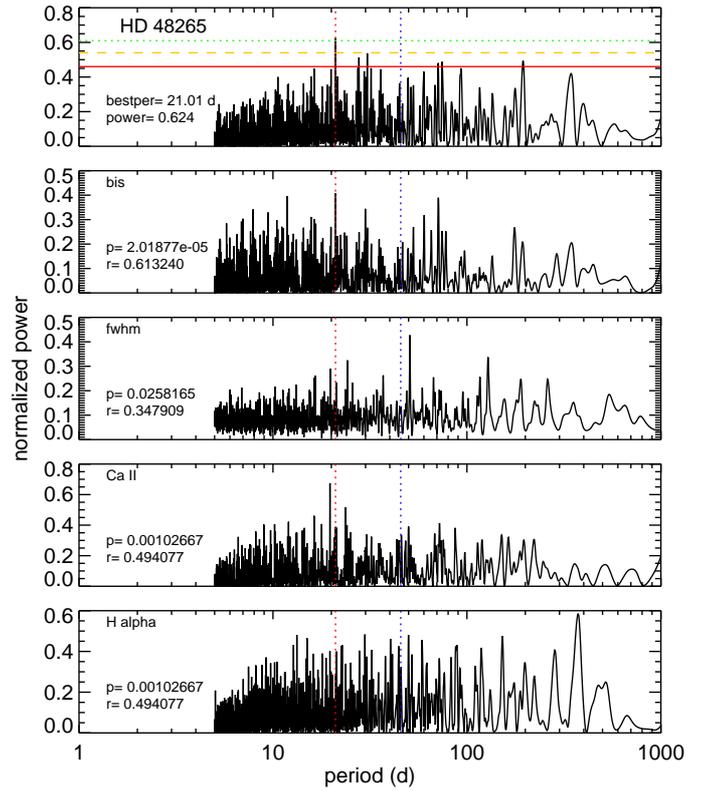}
      \caption{Same as Fig. \ref{fig:hd38801-activity} but for star HD 48265.}
      \label{fig:hd48265-activity}
    \end{figure}

 A power peak with FAP=$0.05$\% is found at 21 days in the residual radial velocity data periodogram, the only 
	  stellar activity-related peak at this period being present in the bisector periodogram (see Fig. \ref{fig:hd48265-activity}). 
	  We therefore tried to find a two-planet solutions for the system, using this residual period and a zero value for eccentricy 
	  for the additional planet's initial setting, but found no satisfactory convergence, suggesting that this residual power 
	  should be regarded as having stellar origins.
	
	\paragraph{HD 50499}
	
	  \begin{table} 
	    \small
		  \caption{Fit results comparison for system HD 50499}\label{table:hd50499}
		  \centering
		  \begin{tabular}{l c c c}
		    \hline\hline
		      \multicolumn{4}{c}{HD 50499}\\
		    \hline
						& \cite{vogt2005}	& \multicolumn{2}{c}{This work}\\
		      Parameter			& Planet b		& Planet b			 & Planet c\\
		    \hline
		      K (ms$^{-1}$)		& $22.9\pm3$ 		& $21.99_{-1.02}^{+1.08}$	 & $\geq8.15 $ \\[3pt]
		      P (days)			& $2482.7\pm110$	& $2447.10_{-21.72}^{+21.92}$	 & $\geq8256.51$ \\[3pt]
		      $\sqrt{e}\cos{\omega}$	& - 			& $0.161_{-0.089}^{+0.084}$	 & - \\[3pt]
		      $\sqrt{e}\sin{\omega}$	& - 			& $-0.483_{-0.041}^{+0.046}$	 & - \\[3pt]
		      e				& $0.23\pm0.14$ 	& $0.266_{-0.044}^{+0.044}$	 & - \\[3pt]
		      $\omega$ (deg)		& $262\pm36$ 		& $288.31_{-10.094}^{+9.579}$	 & - \\[3pt]
		      M$\sin{i}$ (M$_J$)	& $1.71\pm0.2$ 		& $1.636_{-0.017}^{+0.017}$	 & $\geq0.942 $ \\[3pt]
		      a (AU)			& $3.86\pm0.6$ 		& $3.833_{-0.030}^{+0.0305}$	 & - \\[3pt]
		      T$_{peri}$ (days)		&$2451234.6\pm225$	& $2456291.1_{-57.6}^{+58.7}$	 & - \\[3pt]
		    \hline
			n$_{obs}$		& 86		& \multicolumn{2}{c}{24} \\[3pt]
		    \hline
						& mean RV error	& $\gamma$		& jitter			\\
						& (ms$^{-1}$)	& (ms$^{-1}$)		& (ms$^{-1}$)\\[3pt]
		    \hline
		      Keck			& 1.71		& $2.33_{-1.07}^{+1.05}$ & $5.08_{-0.42}^{+0.47}$  \\[3pt]
		      HARPS			& 0.49		& $-18.45_{-3.55}^{+3.57}$ & $2.15_{-0.31}^{+0.39}$ \\[3pt]
		    \hline
		  \end{tabular}
	  \end{table}
	  \begin{figure}
	    \includegraphics[width=\linewidth]{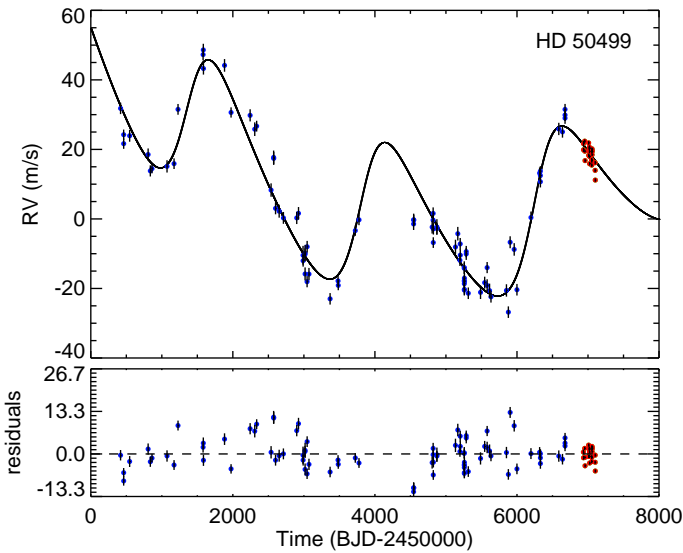}
	    \caption{Same as Fig. \ref{fig:hd4208-fit} but for system HD 50499. The literature Keck data are shown in blue, while our HARPS 
		      survey data are shown in red.}
	    \label{fig:hd50499-fit}
	  \end{figure}
	  
	  The existence of a planet orbiting this G2V star was first reported by \citet{vogt2005} based on 35 HIRES 
	  observations; in the discovery paper a linear trend of -4.8 ms$^{-1}$yr$^{-1}$  was also noted, suggesting the presence of an 
	  outer companion estimated to be located beyond 4 AU and having a minimum mass of at least 2 M$_J$. The nature of this 
	  additional companion was not completely determined, and although an outer star or brown dwarf were dismissed due to 
	  system stability considerations, the proposed two-planet solution was not significantly superior to a model consisting of one 
	  planet and a linear trend.
	  \par However, the recent publication by \citet{butler2017} of 50 new Keck measurements taken over 8 years for HD 50499 shows 
	  this outer trend to be parabolic; Butler's work also provides a new 
	  reduction of Vogt's datapoints, forming the complete Keck dataset we joined with the 24 HARPS data collected during our 
	  observation to refine the orbital solution for HD 50499 b and provide lower limits on the outer planet's mass and period. 
	  Following again the example set by \citet{kipping2011} we fit the radial velocity data for a single Keplerian orbit plus a 
	  quadratic term, obtaining a best-fit curve (see Table \ref{table:hd50499} and Fig. \ref{fig:hd50499-fit}) returning for 
	  planet b values for semiamplitude 
	  $21.99_{-1.02}^{+1.08}$ms$^{-1}$, minimum mass of $1.636_{-0.017}^{+0.017}$M$_J$, period $2447.10_{-21.72}^{+21.92}$d 
	  and eccentricity $0.266_{-0.044}^{+0.044}$, in addition to quadratic coefficients 
	  $k_2$=$(-5.6\pm0.3)\cdot10^{-3}$ms$^{-1}$d$^{-1}$ and $k_2$=$(4.6\pm0.4)\cdot10^{-5}$ms$^{-1}$d$^{-1}$.
	  \par We finally derive 
	  lower limits for the outer planet orbital parameters as P$\geq8256.51$ d, K$\geq8.15$ ms$^{-1}$ and M$\sin{i}\geq0.942$ M$_J$; 
	  we additionally note that \cite{kospal2009} detects an excess in the star's emission at 70 $\mu$m, suggesting the presence of 
	  a debris disk at distances larger than 4-5 AU that can provide further constraints on the outer planet's orbital elements; 
	  from the outer planet's orbital limits it is possible estimate its major semiaxis as lying beyond $\simeq$8.6 AU and 
	  therefore could be placed outside the debris disk. Finally, we find using Hill's criterion the maximum dynamically stable 
	  inner period for any additional inner companion to be 959.5 days.
	  \par The highest power peak in the residual periodogram's is found at about 13 days and has a FAP level of $4$\%, above 
	  our threshold of 1\%.
	
	\paragraph{HD 66428}
	  
	  This G8IV star was first discovered to host a planet in \citet{butler2006} using 22 Keck observations and 
	  its orbital elements were later refined in \citet{feng2015} using 33 datapoints obtained after the upgrade of HIRES, also 
	  reporting a linear trend of $-3.4\pm0.2$ms$^{-1}$yr$^{-1}$ suggesting the presence of an outer companion.
	  \par
\begin{figure}
	    \includegraphics[width=\linewidth]{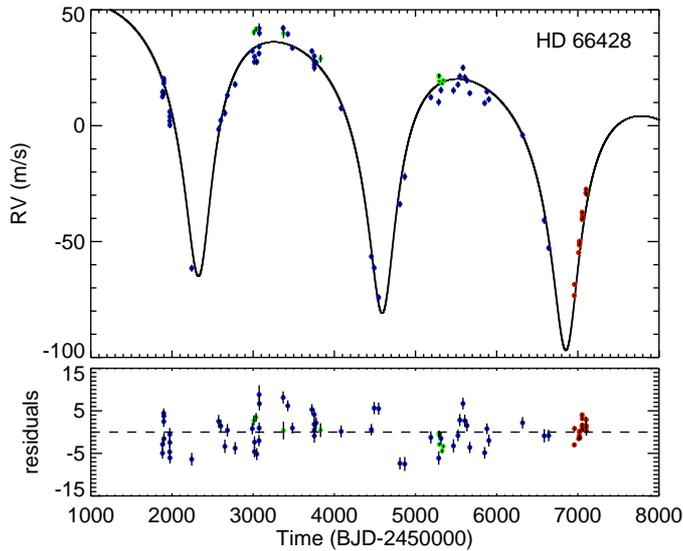}
	    \caption{Same as Fig. \ref{fig:hd4208-fit} but for system HD 66428. The literature Keck data are shown in blue and the literature 
		      HARPS observations are shown in green, while our HARPS survey data are shown in red.}
	    \label{fig:hd66428-fit}
	  \end{figure}

 Combining the improved data reduction published in \citet{butler2017}, 10 HARPS measurements from two different 
	  observation periods publicly available on the ESO 
	  archive and our own 18 HARPS datapoints we found a best-fit solution for HD 66428 b (see Table \ref{table:hd66428} and 
	  Fig. \ref{fig:hd66428-fit}) with 
	  M$\sin{i}$=$3.204_{-0.044}^{+0.043}$M$_J$, P=$2263.12_{-5.94}^{+6.13}$d and e=$0.493_{-0.015}^{+0.015}$, with a linear trend 
	  value of $-2.58_{-0.12}^{+0.11}$ms$^{-1}$yr$^{-1}$. Regarding the system's 
	  linear trend and the possible presence of further bodies in the system, we note that \cite{moutou2017} reports that no 
	  additional outer companion above 0.1 M$_\odot$ has been found in the 0.05 to 50 AU range during a VLT/SPHERE high-angular 
	  resolution survey, and therefore the yet undetected source of the slope in the data must be a substellar companion. 
	  Following the example set by \cite{feng2015}, we can provide an estimate of the minimum mass required of the companion to 
	  produce the detected linear slope $\dot{\gamma}$ over the span of observation $\tau$ using the equation:
	  \begin{equation}\label{eq:slopemass}
	      M_{min}\approx (0.0164\ M_J) \left(\frac{\tau}{yr}\right)^{4/3} \left|\frac{\dot{\gamma}}{ms^{-1}yr^{-1}}\right| 
			     \left(\frac{M_*}{M_\odot}\right)^{2/3}
	  \end{equation}
	  
	  from which we obtain that the outer companion should have a minimum mass of 1.54 M$_J$, a value not dissimilar from the 
	  1.77 M$_J$ value suggested in the same paper. The maximum period allowing dynamical stability for an additional inner planet 
	  is found to be 374.6 days.
	  \par No peak with false alarm probability below 0.01 was found in the residual data periodogram.
	  
	  \begin{table} 
	    \small
		  \caption{Fit results comparison for system HD 66428}\label{table:hd66428}
		  \centering
		  \begin{tabular}{l c c c}
		    \hline\hline
		      \multicolumn{4}{c}{HD 66428}\\
		    \hline
						& \multicolumn{2}{c}{\cite{feng2015}}	& This work\\
		      Parameter			& \multicolumn{2}{c}{Planet b}		& Planet b\\
		    \hline
		      K (ms$^{-1}$)		& \multicolumn{2}{c}{$52.6\pm1.1$} 	& $54.03_{-1.43}^{+1.46}$ \\[3pt]
		      P (days)			& \multicolumn{2}{c}{$2293.9\pm6.4$}	& $2263.12_{-5.94}^{+6.13}$ \\[3pt]
		      $\sqrt{e}\cos{\omega}$	& \multicolumn{2}{c}{-} 		& $-0.696_{-0.010}^{+0.010}$ \\[3pt]
		      $\sqrt{e}\sin{\omega}$	& \multicolumn{2}{c}{-} 		& $-0.086_{-0.023}^{+0.025}$ \\[3pt]
		      e				& \multicolumn{2}{c}{$0.440\pm0.013$}	& $0.493_{-0.015}^{+0.015}$ \\[3pt]
		      $\omega$ (deg)		& \multicolumn{2}{c}{$180.4\pm2.6$} 	& $187.03_{-2.07}^{+1.926}$ \\[3pt]
		      M$\sin{i}$ (M$_J$)	& \multicolumn{2}{c}{$3.194\pm0.060$}	& $3.204_{-0.043}^{+0.043}$ \\[3pt]
		      a (AU)			& \multicolumn{2}{c}{$3.471\pm0.069$}	& $3.467_{-0.024}^{+0.024}$ \\[3pt]
		      T$_{peri}$ (days)		& \multicolumn{2}{c}{$2452278\pm16$} 	& $2456857.2_{-7.4}^{+7.3}$ \\[3pt]
		      slope (ms$^{-1}$yr$^{-1}$)& \multicolumn{2}{c}{$-3.4\pm0.2$}	& $-2.58_{-0.12}^{+0.12}$\\[3pt]
		    \hline
			n$_{obs}$		& \multicolumn{2}{c}{55}		&  83\\[3pt]
		    \hline
						& mean RV error	& $\gamma$		& jitter			\\
						& (ms$^{-1}$)	& (ms$^{-1}$)		& (ms$^{-1}$)\\[3pt]
		    \hline
		      Keck			& 1.40		& $-15.18_{-0.64}^{+0.64}$ & $4.00_{-0.47}^{+0.55}$ \\[3pt]
		      HARPS			& 0.68		& $18.83_{-1.01}^{+1.10}$ & $2.21_{-0.39}^{+0.45}$ \\[3pt]
		    \hline
		  \end{tabular}
	  \end{table}
	  
	  \begin{table} 
	    \small
		  \caption{Fit results comparison for system HD 70642}\label{table:hd70642}
		  \centering
		  \begin{tabular}{l c c c}
		    \hline\hline
		      \multicolumn{4}{c}{HD 70642}\\
		    \hline
						& \multicolumn{2}{c}{\cite{butler2006}}	& This work\\
		      Parameter			& \multicolumn{2}{c}{Planet b}		& Planet b\\
		    \hline
		      K (ms$^{-1}$)		& \multicolumn{2}{c}{$30.4\pm1.3$} 	& $30.40_{-1.91}^{+1.83}$ \\[3pt]
		      P (days)			& \multicolumn{2}{c}{$2068\pm39$} 	& $2124.54_{-13.51}^{+14.65}$ \\[3pt]
		      $\sqrt{e}\cos{\omega}$	& \multicolumn{2}{c}{-}			& $0.020_{-0.102}^{+0.108}$ \\[3pt]
		      $\sqrt{e}\sin{\omega}$	& \multicolumn{2}{c}{-}			& $-0.405_{-0.054}^{+0.059}$ \\[3pt]
		      e				& \multicolumn{2}{c}{$0.034\pm0.043$}	& $0.175_{-0.044}^{+0.044}$ \\[3pt]
		      $\omega$ (deg)		& \multicolumn{2}{c}{$205$} 		& $272.840_{-14.002}^{+15.584}$ \\[3pt]
		      M$\sin{i}$ (M$_J$)	& \multicolumn{2}{c}{$1.97\pm0.18$} 	& $1.993_{-0.018}^{+0.018}$ \\[3pt]
		      a (AU)			& \multicolumn{2}{c}{$3.23\pm0.19$} 	& $3.318_{-0.021}^{+0.022}$ \\[3pt]
		      T$_{peri}$ (days)		& \multicolumn{2}{c}{$2451350\pm380$} 	& $2456159.1_{-131.9}^{+1912.3}$ \\[3pt]
		    \hline
			n$_{obs}$		& \multicolumn{2}{c}{21}		&  50\\[3pt]
		    \hline
						& mean RV error	& $\gamma$		& jitter			\\
						& (ms$^{-1}$)	& (ms$^{-1}$)		& (ms$^{-1}$)\\[3pt]
		    \hline
		      AAT			& 0.75		& $-6.71_{-1.25}^{+1.22}$ & $3.99_{-1.26}^{+1.46}$ \\[3pt]
		      HARPS			& 0.38		& $5.00_{-2.25}^{+2.08}$ & $2.24_{-0.31}^{+0.38}$ \\[3pt]
		    \hline
		  \end{tabular}
	  \end{table}
	  \begin{figure}[]
	    \includegraphics[width=\linewidth]{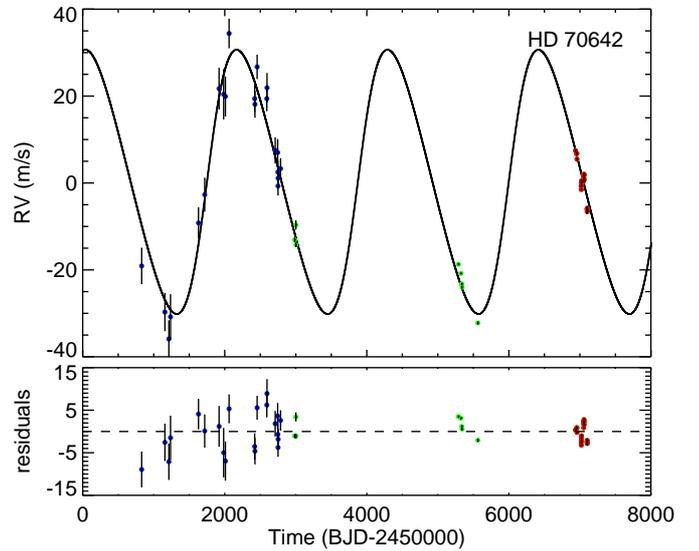}
	    \caption{Same as Fig. \ref{fig:hd4208-fit} but for system HD 70642. The literature AAT data are shown in blue, the literature 
		      HARPS data are shown in green while our HARPS survey data are shown in red.}
	    \label{fig:hd70642-fit}
	  \end{figure}
	  
	\paragraph{HD 70642}
	  
	  A G5V star, its known planet was discovered by \citet{carter2003} using 21 AAT measurements obtained over 
	  5 years of observations; a more tightly constrained orbital solution was later found in \citet{butler2006}.
	  \par In addition to 8 archival HARPS datapoints, the acquisition of our 21 high-precision HARPS measurements 
	  however lead us to a different orbital solution. We find the best-fit curve (see Table \ref{table:hd70642} and 
	  Fig. \ref{fig:hd70642-fit}) to be a Keplerian with semiaplitude 
	  K=$30.40_{-1.91}^{+1.83}$ms$^{-1}$ and period P=$2124.54_{-13.51}^{+14.65}$d; from this we characterize HD 70642 b as having 
	  minimum mass M$\sin{i}$=$1.99_{-0.018}^{+0.018}$M$_J$, eccentricity e=$0.175_{-0.044}^{+0.045}$ and periastron longitude 
	  $\omega$=$272.840_{-14.002}^{+15.584}$deg; the differences in our fit are especially evident in the values of P and $\omega$, 
	  and we also note that our solution features a poorly constrained value for the time of periastron 
	  T$_{peri}$=$2456159.1_{-131.9}^{+1912.3}$d. The maximum period ensuring dynamical stability for an additional inner planet is 
	  estimated via Hill's criterion to be 1232.9 days.
	  \par No residual velocity periodogram peak with a FAP lower than 1\% is found.
	  
	\paragraph{HD 73267}
	  
	  The planet hosted by this G5V star was discovered using 39 HARPS observations as detailed in 
	  \citet{moutou2009}; 10 additional archival HARPS data can also be found on the ESO archive. We exclude from the following 
	  analysis five literature datapoints taken on epochs 2453781.72, 2454121.74, 2454232.56, 2454257.49 and 2455563.60 for having 
	  low SNR.
	  \par
\begin{figure}
	    \includegraphics[width=\linewidth]{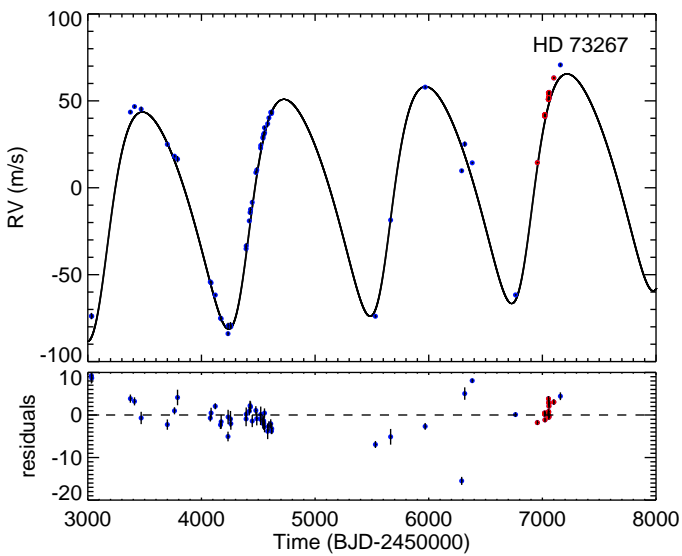}
	    \caption{Same as Fig. \ref{fig:hd4208-fit} but for system HD 73267. The literature HARPS datapoints are shown in blue 
		      while our HARPS survey data are shown in red.}
	    \label{fig:hd73267-fit}
	  \end{figure}

 After joining this literature data with our 17 new HARPS datapoints we found the best-fit curve 
	  (see Table \ref{table:hd73267} and Fig. \ref{fig:hd73267-fit}) to have a semiamplitude value of 
	  $64.65_{-0.87}^{+0.86}$ms$^{-1}$ and period $1245.36_{-2.81}^{+2.81}$d, from wich values of mass 
	  $3.097_{-0.043}^{+0.044}$M$_J$ and eccentricity $0.230_{-0.014}^{+0.014}$ can be inferred, in good agreement with the 
	  literature orbital elements. We also find a previously undetected linear trend of $2.14_{-0.19}^{+0.20}$ms$^{-1}$yr$^{-1}$, 
	  from which using Eq. \ref{eq:slopemass} we derive a minimum mass of $0.83$M$_J$ for the outer companion. A solution without 
	  linear trend was also tested, returning a Bayesian Information Criterion value of 325.25, higher the the BIC value of 
	  258.15 returned by the fit featuring a slope, which is therefore preferred.
	  The dynamical stability limit for inner planets is found with Hill's criterion to be 416.5 days.	  
	  \par The residual radial velocity periodogram features no peak with a false alarm probability low enough to justify a MCMC 
	  search for additional planetary bodies.
	  \begin{table} 
	    \small
		  \caption{Fit results comparison for system HD 73267}\label{table:hd73267}
		  \centering
		  \begin{tabular}{l c c c}
		    \hline\hline
		      \multicolumn{4}{c}{HD 73267}\\
		    \hline
						& \multicolumn{2}{c}{\cite{moutou2009}}	& This work\\
		      Parameter			& \multicolumn{2}{c}{Planet b}		& Planet b\\
		    \hline
		      K (ms$^{-1}$)		& \multicolumn{2}{c}{$64.29\pm0.48$}	& $64.65_{-0.87}^{+0.86}$ \\[3pt]
		      P (days)			& \multicolumn{2}{c}{$1260\pm7$} 	& $1245.36_{-2.81}^{+2.81}$ \\[3pt]
		      $\sqrt{e}\cos{\omega}$	& \multicolumn{2}{c}{-} 		& $-0.296_{-0.022}^{+0.023}$ \\[3pt]
		      $\sqrt{e}\sin{\omega}$	& \multicolumn{2}{c}{-} 		& $-0.376_{-0.020}^{+0.021}$ \\[3pt]
		      e				& \multicolumn{2}{c}{$0.256\pm0.009$}	& $0.230_{-0.014}^{+0.014}$ \\[3pt]
		      $\omega$ (deg)		& \multicolumn{2}{c}{$229.1\pm1.8$}	& $231.799_{-3.286}^{+3.136}$ \\[3pt]
		      M$\sin{i}$ (M$_J$)	& \multicolumn{2}{c}{$3.06\pm0.07$}	& $3.097_{-0.043}^{+0.044}$ \\[3pt]
		      a (AU)			& \multicolumn{2}{c}{$2.198\pm0.025$}	& $2.187_{-0.016}^{+0.016}$ \\[3pt]
		      T$_{peri}$ (days)		& \multicolumn{2}{c}{$2451821.7\pm16$}	& $2456842.7_{-12.0}^{+11.4}$ \\[3pt]
		      slope (ms$^{-1}$yr$^{-1}$)& \multicolumn{2}{c}{-}			& $2.14_{-0.19}^{+0.21}$\\[3pt]
		    \hline
			n$_{obs}$		& \multicolumn{2}{c}{39}		&  63\\[3pt]
		    \hline
						& mean RV error	& $\gamma$		& jitter			\\
						& (ms$^{-1}$)	& (ms$^{-1}$)		& (ms$^{-1}$)\\[3pt]
		    \hline
		      HARPS			& 1.05		& $-9.12_{-0.71}^{+0.71}$ & $3.70_{-0.35}^{+0.43}$ \\[3pt]
		    \hline
		  \end{tabular}
	  \end{table}
	 
	\paragraph{HD 114729}
    
	  This G0V star's planetary companion orbital elements were first announced and then refined by 
	  \citet{butler2003,butler2006}, both results being obtained using Keck data.
	  \par
\begin{figure}
	    \includegraphics[width=\linewidth]{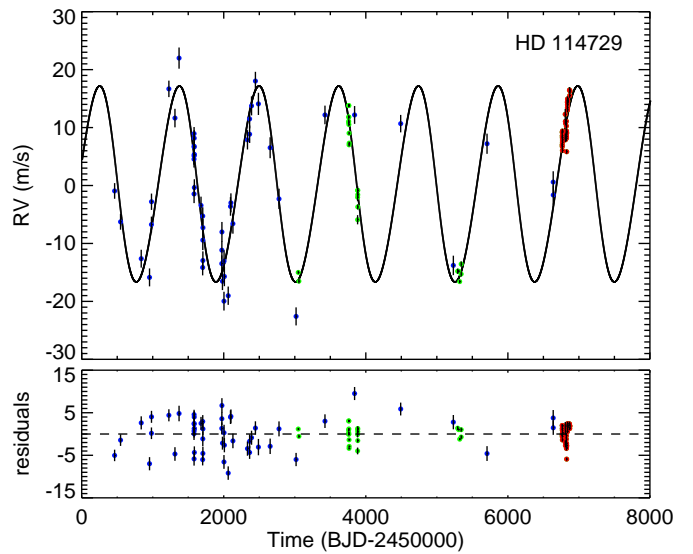}
	    \caption{Same as Fig. \ref{fig:hd4208-fit} but for system HD 114729. The literature Keck data are shown in blue, the literature 
		      HARPS data in green, while our HARPS survey data are shown in red.}
	    \label{fig:hd114729-fit}
	  \end{figure}

 Using both the 34 HARPS data collected during our observations and the improvement in Keck data reduction provided in 
	  \citet{butler2017} in addition to 23 archival HARPS datapoint from the ESO Archive we obtained a similar yet better 
	  constrained orbital fit (see Table \ref{table:hd114729} and Fig. \ref{fig:hd114729-fit}), producing semiamplitude 
	  $16.91_{-0.52}^{+0.53}$ms$^{-1}$, period of $1121.79_{-3.43}^{+3.53}$ days, minimum mass $0.825_{-0.007}^{+0.007}$M$_J$ 
	  and a lower eccentricity of $0.0797_{-0.032}^{+0.029}$. Hill's criterion gives us a maximum period of 697.5 days for 
	  dynamical stability of inner planets.
	  \par
\begin{figure}
      \includegraphics[width=\linewidth]{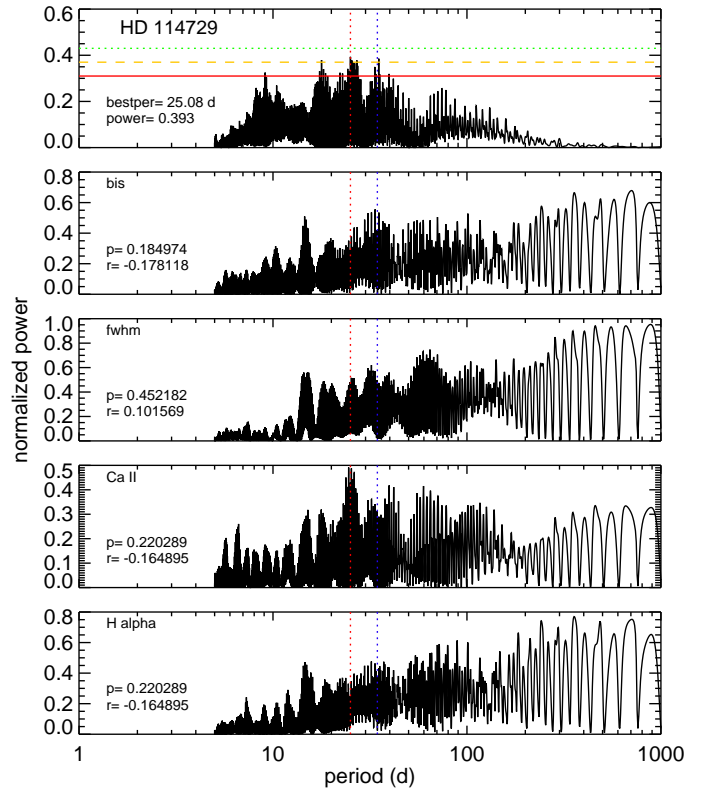}
      \caption{Same as Fig. \ref{fig:hd38801-activity} but for star HD 114729.}
      \label{fig:hd114729-activity}
    \end{figure}

 A major residual periodogram peak at 25 days has a low FAP level of $0.4$\%; this periodogram peak is however 
	  matched by peaks at a similar period in the FWHM and Ca II periodograms (see Fig. \ref{fig:hd114729-activity}), 
	  suggesting a stellar origin for this power peak.
	  
	  \begin{table} 
	    \small
		  \caption{Fit results comparison for system HD 114729}\label{table:hd114729}
		  \centering
		  \begin{tabular}{l c c c}
		    \hline\hline
		      \multicolumn{4}{c}{HD 114729}\\
		    \hline
						& \multicolumn{2}{c}{\cite{butler2006}}	& This work\\
		      Parameter			& \multicolumn{2}{c}{Planet b}		& Planet b\\
		    \hline
		      K (ms$^{-1}$)		& \multicolumn{2}{c}{$18.8\pm1.3$}	& $16.91_{-0.52}^{+0.53}$ \\[3pt]
		      P (days)			& \multicolumn{2}{c}{$1114\pm15$}	& $1121.79_{-3.43}^{+3.53}$ \\[3pt]
		      $\sqrt{e}\cos{\omega}$	& \multicolumn{2}{c}{-}			& $0.061_{-0.124}^{+0.112}$ \\[3pt]
		      $\sqrt{e}\sin{\omega}$	& \multicolumn{2}{c}{-}			& $0.251_{-0.086}^{+0.058}$ \\[3pt]
		      e				& \multicolumn{2}{c}{$0.167\pm0.055$} 	& $0.079_{-0.032}^{+0.029}$ \\[3pt]
		      $\omega$ (deg)		& \multicolumn{2}{c}{$93\pm30$}		& $77.059_{-27.685}^{+27.681}$ \\[3pt]
		      M$\sin{i}$ (M$_J$)	& \multicolumn{2}{c}{$0.95\pm0.10$} 	& $0.825_{-0.007}^{+0.007}$ \\[3pt]
		      a (AU)			& \multicolumn{2}{c}{$2.11\pm0.12$} 	& $2.067_{-0.010}^{+0.010}$ \\[3pt]
		      T$_{peri}$ (days)		& \multicolumn{2}{c}{$2450520\pm67$}	& $2454947.7_{-82.8}^{+82.4}$ \\[3pt]
		    \hline
			n$_{obs}$		& \multicolumn{2}{c}{52}		& 109 \\[3pt]
		    \hline
						& mean RV error	& $\gamma$		& jitter			\\
						& (ms$^{-1}$)	& (ms$^{-1}$)		& (ms$^{-1}$)\\[3pt]
		    \hline
		      Keck			& 1.14		& $1.46_{-0.63}^{+0.64}$ & $3.93_{-0.44}^{+0.50}$ \\[3pt]
		      HARPS			& 0.22		& $-6.11_{-0.42}^{+0.41}$ & $2.00_{-0.19}^{+0.22}$ \\[3pt]
		    \hline
		  \end{tabular}
	  \end{table}
	 
	\paragraph{HD 117207}
	  
	  A G8IV star, its planet was first announced in \citet{marcy2005} and the orbital solution was later 
	  refined by \citet{butler2006}. This literature data were later refined and newly reduced in \citet{butler2017}.
	  \par
\begin{figure}
	    \includegraphics[width=\linewidth]{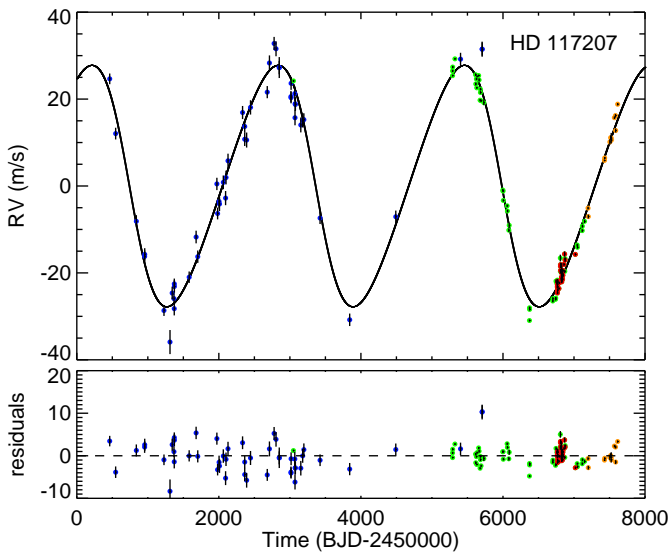}
	    \caption{Same as Fig. \ref{fig:hd4208-fit} but for system HD 117207. The literature Keck data are shown in blue, past HARPS data 
		      in green, our HARPS survey data are shown in red while literature HARPS data taken after the fiber-link update of 
		      May 2015 are in orange.}
	    \label{fig:hd117207-fit}
	  \end{figure}

 We study the timeseries resulting in joining this newly reduced Keck data, 44 archival HARPS measurements, our own 33 
	  HARPS datapoints and 12 further archival HARPS measurements taken after the May 2015 fiber-link update and that are here 
	  therefore treated as a independent dataset. We obtain a a significantly better constrained orbital solution 
	  (see Table \ref{table:hd117207} and Fig. \ref{fig:hd117207-fit}), with minimum mass $1.926_{-0.034}^{+0.034}$M$_J$, period 
	  of $2621.75_{-8.53}^{+8.37}$d, eccentricity $0.157_{-0.013}^{+0.013}$ as well as periastron longitude of 
	  $90.217_{-5.523}^{+5.275}$deg. The maximum period allowing dynamical stability for an inner planet is found with Hill's 
	  criterion to be 1263.5 days.
	  \par No power peak with FAP$\leq0.01$ is found in the residual data periodogram.
	  
	  \begin{table} 
	    \small
		  \caption{Fit results comparison for system HD 117207}\label{table:hd117207}
		  \centering
		  \begin{tabular}{l c c c}
		    \hline\hline
		      \multicolumn{4}{c}{HD 117207}\\
		    \hline
						& \multicolumn{2}{c}{\cite{butler2006}}	& This work\\
		      Parameter			& \multicolumn{2}{c}{Planet b}		& Planet b \\[3pt]
		    \hline
		      K (ms$^{-1}$)		& \multicolumn{2}{c}{$26.6\pm0.93$} 	& $27.78_{-0.33}^{+0.33}$ \\[3pt]
		      P (days)			& \multicolumn{2}{c}{$2597\pm41$} 	& $2621.75_{-8.53}^{+8.37}$ \\[3pt]
		      $\sqrt{e}\cos{\omega}$	& \multicolumn{2}{c}{-} 		& $-0.001_{-0.037}^{+0.038}$ \\[3pt]
		      $\sqrt{e}\sin{\omega}$	& \multicolumn{2}{c}{-} 		& $0.394_{-0.017}^{+0.016}$ \\[3pt]
		      e				& \multicolumn{2}{c}{$0.144\pm0.035$} 	& $0.157_{-0.013}^{+0.013}$ \\[3pt]
		      $\omega$ (deg)		& \multicolumn{2}{c}{$73\pm16$} 	& $90.217_{-5.523}^{+5.275}$ \\[3pt]
		      M$\sin{i}$ (M$_J$)	& \multicolumn{2}{c}{$1.88\pm0.17$} 	& $1.926_{-0.034}^{+0.034}$ \\[3pt]
		      a (AU)			& \multicolumn{2}{c}{$3.79\pm0.22$} 	& $3.787_{-0.035}^{+0.034}$ \\[3pt]
		      T$_{peri}$ (days)		& \multicolumn{2}{c}{$2450630\pm120$} 	& $2455978.9_{-37.8}^{+35.7}$ \\[3pt]
		    \hline
			n$_{obs}$		& \multicolumn{2}{c}{51}		& 140 \\[3pt]
		    \hline
						& mean RV error	& $\gamma$		& jitter			\\
						& (ms$^{-1}$)	& (ms$^{-1}$)		& (ms$^{-1}$)\\[3pt]
		    \hline
		      Keck			& 1.54		& $-10.58_{-0.53}^{+0.52}$& $3.44_{-0.40}^{+0.47}$ \\[3pt]
		      HARPS1			& 0.35		& $4.65_{-0.29}^{+0.29}$ & $1.71_{-0.14}^{+0.16}$ \\[3pt]
		      HARPS2			& 0.42		& $21.38_{-0.84}^{+0.85}$ & $1.88_{-0.38}^{+0.55}$ \\[3pt]
		    \hline
		  \end{tabular}
	  \end{table}
	  
	\paragraph{HD 126525}
      
	  This G4V star has been target to various HARPS observational surveys, and \citet{mayor2011} summary work 
	  reports a best-fit solution for its planet; no value for time of periastron passage or longitude is given in the discovery 
	  paper, nor the number of measurements used to obtain the orbital solution.
	  \par We collected 35 HARPS measurements in our observations, from which we removed a single datapoint at epoch $2456869.5$ 
	  due to low SNR value. A total of 96 archival HARPS datapoints were also collected, 11 of which were taken after the 
	  fiber-link update of May 2015 and are here treated as an independent dataset.
	  \par
\begin{figure}
	    \includegraphics[width=\linewidth]{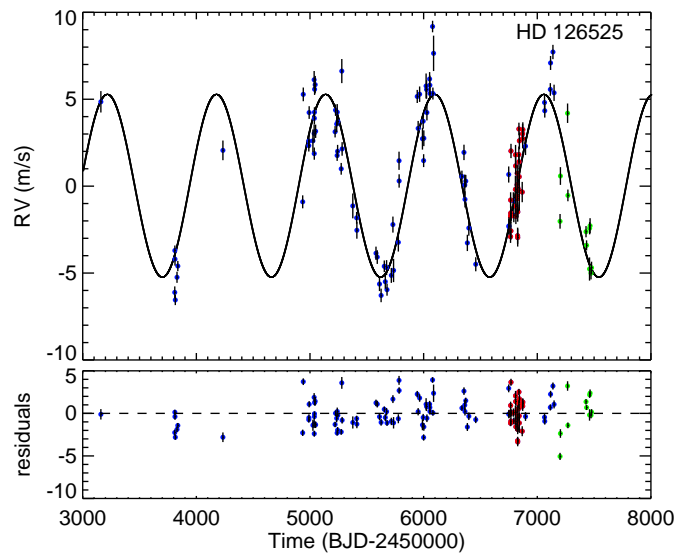}
	    \caption{Same as Fig. \ref{fig:hd4208-fit} but for system HD 126525. Past literature HARPS data are shown in blue, 
		      while our HARPS survey data are shown in red and literature HARPS datapoints taken afet the fiber-link update of May 
		      2015 are shown in green.}
	    \label{fig:hd126525-fit}
	  \end{figure}

 Joining the literature and the new data we find a fit (see Table \ref{table:hd126525} and Fig. 
	  \ref{fig:hd126525-fit}) in which the planet has a minimum mass of $0.237_{-0.002}^{+0.002}$M$_J$, a period of 
	  $960.41_{-6.33}^{+6.19}$d and an eccentricity value of $0.035_{-0.024}^{+0.039}$, a solution generally better constrained 
	  than the literature one. The limit on dynamical stability for an additional inner companion is estimated with Hill's 
	  criterion to be 707.7 days.
	  \par No significant peak is found in the residual data periodogram.
	 
	  \begin{table} 
	    \small
		  \caption{Fit results comparison for system HD 126525}\label{table:hd126525}
		  \centering
		  \begin{tabular}{l c c c}
		    \hline\hline
		      \multicolumn{4}{c}{HD 126525}\\
		    \hline
						& \multicolumn{2}{c}{\cite{mayor2011}}	& This work\\
		      Parameter			& \multicolumn{2}{c}{Planet b}		& Planet b\\
		    \hline
		      K (ms$^{-1}$)		& \multicolumn{2}{c}{$5.11\pm0.34$} 	& $5.26_{-0.25}^{+0.25}$ \\[3pt]
		      P (days)			& \multicolumn{2}{c}{$948.1\pm22$} 	& $960.41_{-6.33}^{+6.19}$ \\[3pt]
		      $\sqrt{e}\cos{\omega}$	& \multicolumn{2}{c}{-} 		& $0.045_{-0.151}^{+0.143}$ \\[3pt]
		      $\sqrt{e}\sin{\omega}$	& \multicolumn{2}{c}{-} 		& $-0.071_{-0.130}^{+0.151}$ \\[3pt]
		      e				& \multicolumn{2}{c}{$0.13\pm0.07$} 	& $0.035_{-0.024}^{+0.039}$ \\[3pt]
		      $\omega$ (deg)		& \multicolumn{2}{c}{-} 		& $247.506_{-184.079}^{+76.249}$ \\[3pt]
		      M$\sin{i}$ (M$_J$)	& \multicolumn{2}{c}{$0.224\pm0.0182$} 	& $0.237_{-0.002}^{+0.002}$ \\[3pt]
		      a (AU)			& \multicolumn{2}{c}{$1.811\pm0.041$} 	& $1.837_{-0.010}^{+0.010}$ \\[3pt]
		      T$_{peri}$ (days)		& \multicolumn{2}{c}{-} 		& $2456161.7_{-215.7}^{+534.8}$ \\[3pt]
		    \hline
			n$_{obs}$		& \multicolumn{2}{c}{-}		& 130 \\[3pt]
		    \hline
						& mean RV error	& $\gamma$		& jitter			\\
						& (ms$^{-1}$)	& (ms$^{-1}$)		& (ms$^{-1}$)\\[3pt]
		    \hline
		      HARPS1			& 0.45		& $-2.18_{-0.18}^{+0.18}$ & $1.56_{-0.10}^{+0.12}$ \\[3pt]
		      HARPS2			& 0.41		& $16.89_{-0.80}^{+0.81}$ & $2.50_{-0.54}^{+0.78}$ \\[3pt]
		    \hline
		  \end{tabular}
	  \end{table}
	
	\paragraph{HD 143361}
	
	  \begin{table} 
	    \small
		  \caption{Fit results comparison for system HD 143361}\label{table:hd143361}
		  \centering
		  \begin{tabular}{l c c c}
		    \hline\hline
		      \multicolumn{4}{c}{HD 143361}\\
		    \hline
						& \multicolumn{2}{c}{\cite{jenkins2017}}& This work\\
		      Parameter			& \multicolumn{2}{c}{Planet b}		& Planet b\\
		    \hline
		      K (ms$^{-1}$)		& \multicolumn{2}{c}{$72.1\pm1$}	& $73.89_{-0.58}^{+0.56}$ \\[3pt]
		      P (days)			& \multicolumn{2}{c}{$1046.2\pm3.2$} 	& $1039.15_{-1.70}^{+1.64}$ \\[3pt]
		      $\sqrt{e}\cos{\omega}$	& \multicolumn{2}{c}{-}			& $-0.209_{-0.014}^{+0.015}$ \\[3pt]
		      $\sqrt{e}\sin{\omega}$	& \multicolumn{2}{c}{-}			& $-0.392_{-0.009}^{+0.010}$ \\[3pt]
		      e				& \multicolumn{2}{c}{$0.193\pm0.015$} 	& $0.197_{-0.006}^{+0.006}$ \\[3pt]
		      $\omega$ (deg)		& \multicolumn{2}{c}{$241.22\pm3.44$} 	& $241.889_{-2.051}^{+2.112}$ \\[3pt]
		      M$\sin{i}$ (M$_J$)	& \multicolumn{2}{c}{$3.48\pm0.24$} 	& $3.532_{-0.066}^{+0.065}$ \\[3pt]
		      a (AU)			& \multicolumn{2}{c}{$1.98\pm0.07$} 	& $1.988_{-0.018}^{+0.018}$ \\[3pt]
		      T$_{peri}$ (days)		& \multicolumn{2}{c}{$2453746\pm147$} 	& $2456805.8_{-4.9}^{+4.9}$ \\[3pt]
		    \hline
			n$_{obs}$		& \multicolumn{2}{c}{79}		& 107 \\[3pt]
		    \hline
						& mean RV error	& $\gamma$		& jitter			\\
						& (ms$^{-1}$)	& (ms$^{-1}$)		& (ms$^{-1}$)\\[3pt]
		    \hline
		      MIKE			& 3.13		& $7.50_{-3.75}^{+3.87}$ & $11.47_{-2.49}^{+3.59}$ \\[3pt]
		      HARPS			& 0.87		& $10.15_{-0.48}^{+0.48}$ & $1.86_{-0.22}^{+0.25}$ \\[3pt]
		      CORALIE			& 14.42		& $3.31_{-2.11}^{+2.09}$ & $2.23_{-1.57}^{+2.47}$ \\[3pt]
		    \hline
		  \end{tabular}
	  \end{table}
	  \begin{figure}
	    \includegraphics[width=\linewidth]{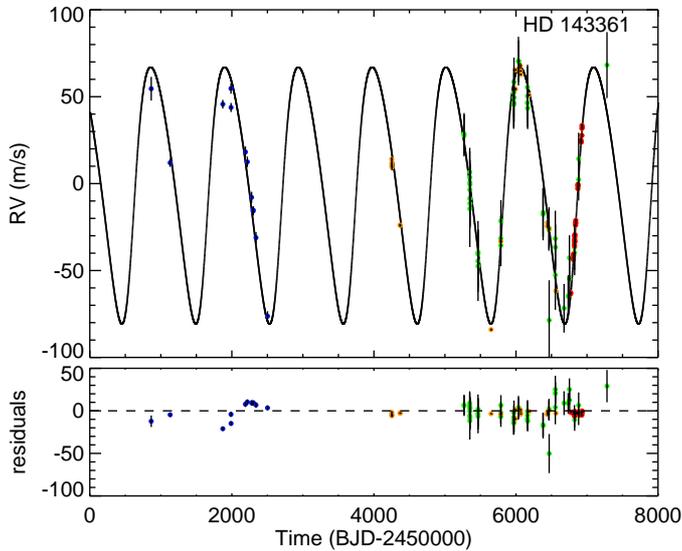}
	    \caption{Same as Fig. \ref{fig:hd4208-fit} but for system HD 143361. The literature MIKE data are shown in blue, literature 
		      HARPS data in orange, literature CORALIE datapoints are marked in green, while our HARPS survey data are shown in red.}
	    \label{fig:hd143361-fit}
	  \end{figure}
    
	  The planet orbiting this G6V star was first discovered and reported in \citet{minniti2009} as the
	  result of 12 MIKE observation, and later studies \citep{jenkins2009,jenkins2017} refined the known orbit based on additional 
	  HARPS and CORALIE datapoints.
	  \par We joined literature data, excluding from the 22 HARPS datapoints the three low-SNR measurements taken at epochs 
	  2454253.77, 2454578.78 and 2454581.80, with our 32 HARPS measurements, from which we exclude a single point at epoch 
	  2456869.62 due to low SNR value. We thusly obtain a best-fit curve (see Table \ref{table:hd143361} and Fig. 
	  \ref{fig:hd143361-fit}) of semiamplitude $73.89_{-0.58}^{+0.56}$ms$^{-1}$ and a $1039.15_{-1.70}^{+1.64}$ days period, 
	  slightly shorter than the literature result. We then obtain a minimum mass of $3.532_{-0.066}^{+0.065}$M$_J$ and an 
	  eccentricity value of $0.197_{-0.006}^{+0.006}$, a solution that agrees well with the results of \citet{jenkins2017} and yet 
	  is far better constrained.
	  \par The residual data periodogram features no major peak with FAP$\leq0.01$.
	  
	\paragraph{HD 152079}
	
	  \begin{table} 
	    \small
		  \caption{Fit results comparison for system HD 152079}\label{table:hd152079}
		  \centering
		  \begin{tabular}{l c c c}
		    \hline\hline
		      \multicolumn{4}{c}{HD 152079}\\
		    \hline
						& \multicolumn{2}{c}{\cite{jenkins2017}}	& This work\\
		      Parameter			& \multicolumn{2}{c}{Planet b}		& Planet b\\
		    \hline
		      K (ms$^{-1}$)		& \multicolumn{2}{c}{$31.3\pm1.1$}	& $40.76_{-1.10}^{+1.16}$ \\[3pt]
		      P (days)			& \multicolumn{2}{c}{$2899\pm52$} 	& $2918.92_{-39.28}^{+37.87}$ \\[3pt]
		      $\sqrt{e}\cos{\omega}$	& \multicolumn{2}{c}{-} 		& $0.648_{-0.018}^{+0.018}$ \\[3pt]
		      $\sqrt{e}\sin{\omega}$	& \multicolumn{2}{c}{-} 		& $-0.331_{-0.024}^{+0.025}$ \\[3pt]
		      e				& \multicolumn{2}{c}{$0.52\pm0.02$} 	& $0.532_{-0.016}^{+0.015}$ \\[3pt]
		      $\omega$ (deg)		& \multicolumn{2}{c}{$324.87\pm3.44$} 	& $332.905_{-2.195}^{+2.299}$ \\[3pt]
		      M$\sin{i}$ (M$_J$)	& \multicolumn{2}{c}{$2.18\pm0.17$} 	& $2.661_{-0.046}^{+0.046}$ \\[3pt]
		      a (AU)			& \multicolumn{2}{c}{$3.98\pm0.15$} 	& $4.187_{-0.053}^{+0.051}$ \\[3pt]
		      T$_{peri}$ (days)		& \multicolumn{2}{c}{$2453193\pm260$} 	& $2456173.2_{-11.3}^{+11.6}$ \\[3pt]
		      slope (ms$^{-1}$yr$^{-1}$)& \multicolumn{2}{c}{$1.72\pm0.47$}	& $0.35_{-0.24}^{+0.24}$ \\[3pt]
		    \hline
			n$_{obs}$		& \multicolumn{2}{c}{46}		& 66 \\[3pt]
		    \hline
						& mean RV error	& $\gamma$		& jitter			\\
						& (ms$^{-1}$)	& (ms$^{-1}$)		& (ms$^{-1}$)\\[3pt]
		    \hline
		      MIKE			& 2.98		& $-10.34_{-1.61}^{+1.51}$ & $4.08_{-1.22}^{+1.51}$ \\[3pt]
		      HARPS			& 0.83		& $-7.64_{-0.63}^{+0.63}$ & $2.21_{-0.25}^{+0.30}$ \\[3pt]
		    \hline
		  \end{tabular}
	  \end{table}
	  \begin{figure}
	    \includegraphics[width=\linewidth]{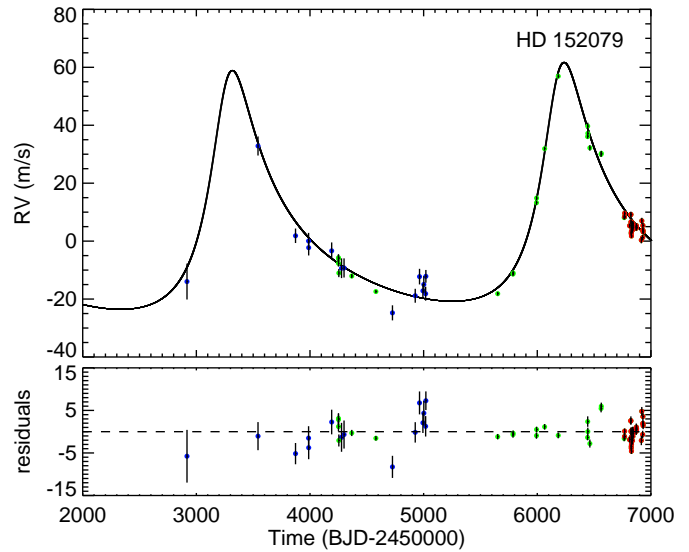}
	    \caption{Same as Fig. \ref{fig:hd4208-fit} but for system HD 152079. Literature MIKE data are shown in blue, literature HARPS 
		      data in green while our HARPS survey data are shown in red.}
	    \label{fig:hd152079-fit}
	  \end{figure}
    
	  This G6V star hosts a planet first reported in \citet{arriagada2010}, a discovery based on 15 MIKE 
	  datapoints, whose orbit was later refined by \citet{jenkins2017} using additional 15 CORALIE measurements and 16 HARPS data 
	  in good agreement with Arriagada's first findings, also reporting a linear trend of $1.72\pm0.47$ms$^{-1}$yr$^{-1}$.
	  \par We joined our 32 datapoints to the literature timeseries (from which two low-SNR datapoints at epochs 2455649.80 and 
	  2455650.75 were excluded) and additional 7 HARPS archival data, obtaining a best-fit (see Table \ref{table:hd152079} and Fig. 
	  \ref{fig:hd152079-fit}) solution having M$\sin{i}$=$2.661_{-0.046}^{+0.046}$M$_J$, period of $2918.92_{-39.28}^{+37.87}$ days 
	  and eccentricity value of $0.532_{-0.016}^{+0.015}$, also featuring a linear trend of 
	  $0.35_{-0.24}^{+0.24}$ms$^{-1}$yr$^{-1}$ which, using again Eq. \ref{eq:slopemass}, suggests the presence of an outer 
	  companion of at least 0.15 M$_J$. While generally agreeing to the previously published fit, we note that our solution 
	  features slightly higher values for planetary mass and major semiaxis and a significantly lower value of linear acceleration. 
	  Hill's criterion gives an estimate of 395.9 days for an inner planet's maximum period allowing dynamical stability.
	  \par
\begin{figure}
      \includegraphics[width=\linewidth]{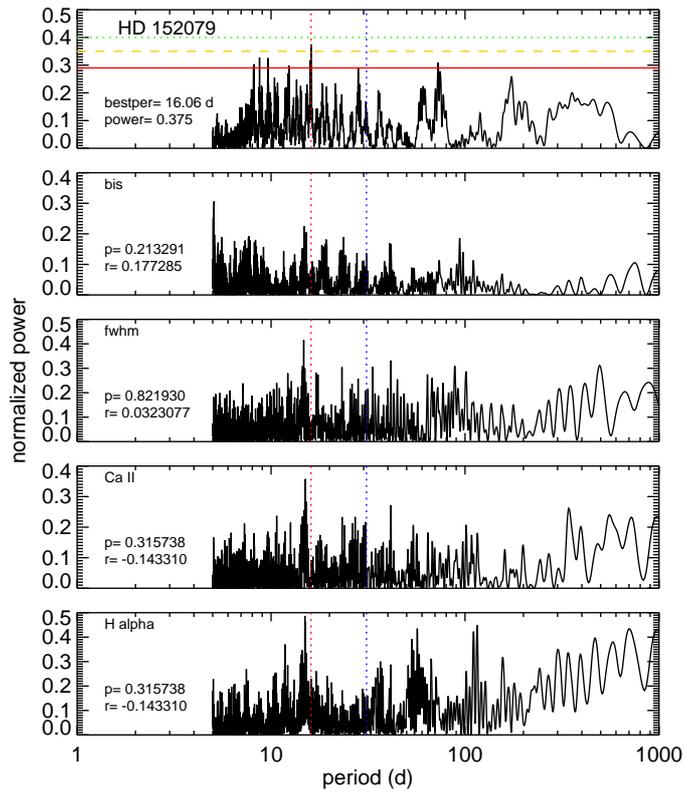}
      \caption{Same as Fig. \ref{fig:hd38801-activity} but for star HD 152079.}
      \label{fig:hd152079-activity}
    \end{figure}

 The residual data periodogram features a major peak at 16 days with a false alarm probability of 0.4\%; however 
	  significant peaks at similar periods can be found in the periodograms obtained for the activity indexes 
	  (see Fig. \ref{fig:hd152079-activity}), suggesting a non-planetary origin for the signal.
	
	\paragraph{HD 187085}
	
	  \begin{table} 
	    \small
		  \caption{Fit results comparison for system HD 187085}\label{table:hd187085}
		  \centering
		  \begin{tabular}{l c c c}
		    \hline\hline
		      \multicolumn{4}{c}{HD 187085}\\
		    \hline
						& \multicolumn{2}{c}{\cite{jones2006}}	& This work\\
		      Parameter			& \multicolumn{2}{c}{Planet b}		& Planet b\\
		    \hline
		      K (ms$^{-1}$)		& \multicolumn{2}{c}{$17$} 	& $15.39_{-1.98}^{+2.21}$ \\[3pt]
		      P (days)			& \multicolumn{2}{c}{$986$} 	& $1019.74_{-22.58}^{+21.29}$ \\[3pt]
		      $\sqrt{e}\cos{\omega}$	& \multicolumn{2}{c}{-} 	& $-0.045_{-0.186}^{+0.165}$ \\[3pt]
		      $\sqrt{e}\sin{\omega}$	& \multicolumn{2}{c}{-} 	& $0.456_{-0.448}^{+0.222}$ \\[3pt]
		      e				& \multicolumn{2}{c}{$0.47$}	& $0.251_{-0.191}^{+0.221}$ \\[3pt]
		      $\omega$ (deg)		& \multicolumn{2}{c}{$94$} 	& $98.315_{-20.027}^{+78.336}$ \\[3pt]
		      M$\sin{i}$ (M$_J$)	& \multicolumn{2}{c}{$0.75$}	& $0.836_{-0.011}^{+0.011}$ \\[3pt]
		      a (AU)			& \multicolumn{2}{c}{$2.05$}	& $2.100_{-0.032}^{+0.032}$ \\[3pt]
		      T$_{peri}$ (days)		& \multicolumn{2}{c}{$2450912$}	& $2456053.6_{-54.4}^{+252.4}$ \\[3pt]
		    \hline
			n$_{obs}$		& \multicolumn{2}{c}{40}	& 69 \\[3pt]
		    \hline
						& mean RV error	& $\gamma$		& jitter	\\
						& (ms$^{-1}$)	& (ms$^{-1}$)		& (ms$^{-1}$)\\[3pt]
		    \hline
		      AAT			& 4.51		& $-1.66_{-1.24}^{+1.20}$ & $5.51_{-1.00}^{+1.15}$ \\[3pt]
		      HARPS			& 0.59		& $-13.50_{-2.00}^{+2.06}$ & $1.82_{-0.27}^{+0.33}$ \\[3pt]
		    \hline
		  \end{tabular}
	  \end{table}
	  \begin{figure}
	    \includegraphics[width=\linewidth]{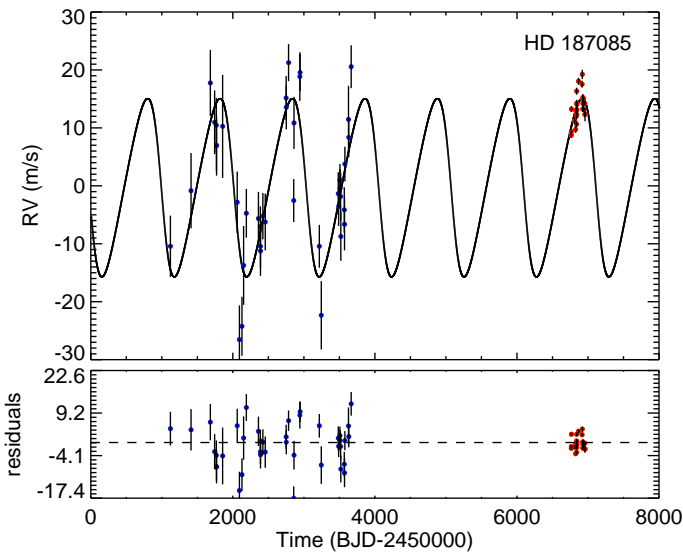}
	    \caption{Same as Fig. \ref{fig:hd4208-fit} but for system HD 187085. The literature AAT data are shown in blue, while our HARPS 
		      survey data are shown in red.}
	    \label{fig:hd187085-fit}
	  \end{figure}
	
	  A G0V star hosting a planet first reported in \citet{jones2006} using 40 AAT measurements; it is however 
	  noted in the discovery paper that the data does not manage to effectively sample the sharp drop evident in the radial 
	  velocities, allowing therefore for a better fit characterized by a lower eccentricity.
	  \par Our 30 HARPS measurements also fail to sample this sharp drop, therefore not conclusively determining whether a lower 
	  value of eccentricity would better fit the data. We also report that the datapoint at epoch $2456825.878$, was taken with a 
	  erroneous exposure time of $5$ seconds and 
	  is therefore ignored in our analysis; the same night (epoch $2456825.884$) a new measurement was immediately taken with a 
	  correct exposure time of 600 seconds, this one being used instead in our analysis.
	  \par We find a new orbital solution (see Table \ref{table:hd187085} and Fig. \ref{fig:hd187085-fit}) with semiaplitude 
	  K=$15.39_{-1.98}^{+2.21}$ms$^{-1}$, 
	  a longer period of $1019.74_{-22.58}^{+21.29}$ days, slightly higher minimum mass of $0.836_{-0.011}^{+0.011}$M$_J$ and 
	  lower but still poorly constrained eccentricity e=$0.251_{-0.191}^{+0.221}$. Although no uncertainty on the published orbital 
	  elements were available in the discovery paper, we note that our solution is generally compatible with the nominal values 
	  provided in \citet{jones2006}. We find the highest orbital period allowing stability for an additional inner planet's orbit to 
	  be 453.2 days.
	  \par The residual periodogram show no low FAP significant peak.
	  
	\paragraph{HD 190647}
	
	  This G5V star has been the target of 20 HARPS observations that, as detailed in \citet{naef2007}, led to 
	  the discovery of planetary companion HD 190647 b. The authors note however that their data failed to cover the entire orbital 
	  period. In the following analysis, we 
	  excluded from this literature timeseries the 4 datapoints taken at epochs 2453273.59, 2453274.60, 2453466.90 and 2453493.92 
	  due to low values of SNR.	  
	  \par
\begin{figure}
	    \includegraphics[width=\linewidth]{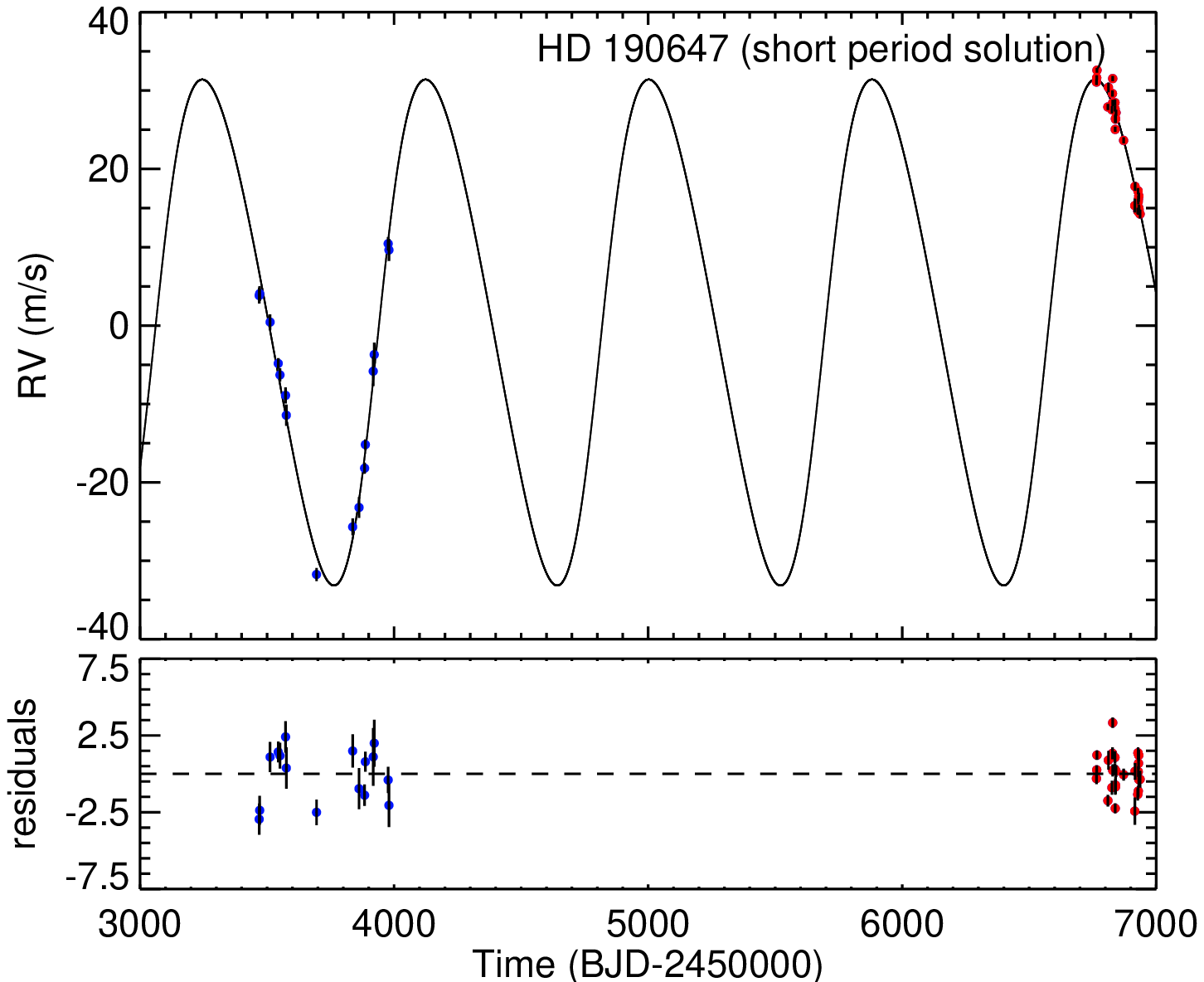}
	    \caption{Same as Fig. \ref{fig:hd4208-fit} but for the short-period solution of system HD 190647. The literature HARPS data are 
		      shown in blue, while our HARPS survey data are shown in red.}
	    \label{fig:hd190647-short-fit}
	  \end{figure}

\begin{figure}
	    \includegraphics[width=\linewidth]{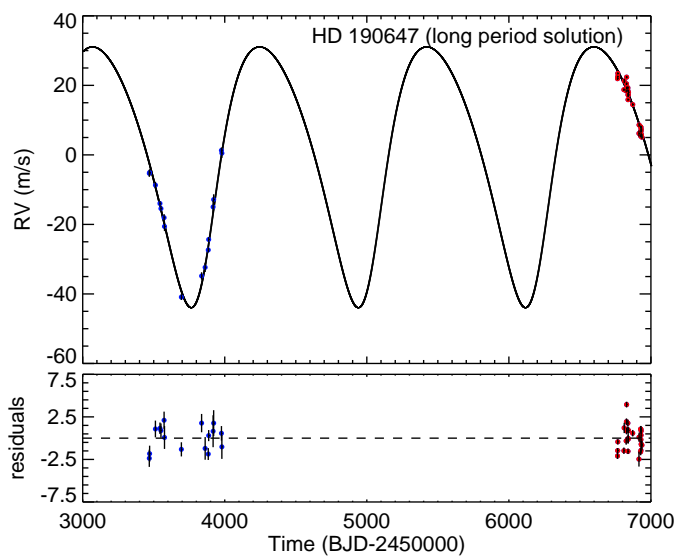}
	    \caption{Same as Fig. \ref{fig:hd4208-fit} but for the long-period solution of system HD 190647. The literature HARPS data are 
		      shown in blue, while our HARPS survey data are shown in red.}
	    \label{fig:hd190647-long-fit}
	  \end{figure}

 The 30 HARPS data collected in our survey however manage to sample more efficiently the maximum in RV variation; we 
	  removed the datapoint at epoch $2456827.918$ since it was taken with an erroneus exposure time of 5 seconds and 
	  another datapoint at epoch 2456840.81 for having low SNR.
	  The resulting timeseries lead to two significantly different possible orbital solutions with similar statistical weight 
	  (see Table \ref{table:hd190647}); since the main difference between these solutions is found in orbital period we shall 
	  refer to them as a 'short period solution' (see Fig. \ref{fig:hd190647-short-fit}) and a 'long period solution' (see Fig. 
	  \ref{fig:hd190647-long-fit}). Having respectively a Bayesian Information Criterion value of 99.79 and 101.21 neither solution 
	  is clearly preferred.
	  \par We characterize the short period solution as having orbital period $878.86_{-1.65}^{+1.76}$d, radial velocity 
	  semiamplitude $32.28_{-0.66}^{+0.68}$ms$^{-1}$, eccentricity of $0.146_{-0.018}^{+0.018}$ and minimum mass of 
	  $1.573_{-0.026}^{+0.026}$M$_J$. The long period solution features instead a period of $1176.45_{-2.83}^{+3.10}$d, 
	  semiamplitude $37.51_{-0.86}^{+0.82}$ms$^{-1}$, eccentricity of $0.224_{-0.014}^{+0.014}$ and minimum planetary mass 
	  $1.985_{-0.033}^{+0.033}$M$_J$.
	  \par No peak with FAP$\leq0.01$ was found in the residual periodogram obtained for either solution. From Hill's 
	  criterion we obtain a stability limit maximum period of 527.4 days for an additional inner planet.
		  
	  \begin{table*}
	    \small
		  \caption{Fit results comparison for system HD 190647}\label{table:hd190647}
		  \centering
		  \begin{tabular}{l c c c c c}
		    \hline\hline
		      \multicolumn{6}{c}{HD 190647}\\
		    \hline
						& 			& \multicolumn{2}{c}{This work}	&\multicolumn{2}{c}{This work}\\
						& \cite{naef2007}	& \multicolumn{2}{c}{(short period solution)}		& \multicolumn{2}{c}{(long period solution)}\\
		      Parameter			& Planet b		& \multicolumn{2}{c}{Planet b}				& \multicolumn{2}{c}{Planet b}\\
		    \hline
		      K (ms$^{-1}$)		& $36.4\pm1.2$		& \multicolumn{2}{c}{$32.28_{-0.66}^{+0.68}$}		& \multicolumn{2}{c}{$37.51_{-0.86}^{+0.82}$} \\[3pt]
		      P (days)			& $1038.1\pm4.9$ 	& \multicolumn{2}{c}{$878.86_{-1.65}^{+1.76}$}		& \multicolumn{2}{c}{$1176.45_{-2.83}^{+3.10}$} \\[3pt]
		      $\sqrt{e}\cos{\omega}$	& - 			& \multicolumn{2}{c}{$-0.071_{-0.041}^{+0.042}$}	& \multicolumn{2}{c}{$-0.364_{-0.033}^{+0.038}$} \\[3pt]
		      $\sqrt{e}\sin{\omega}$	& - 			& \multicolumn{2}{c}{$-0.373_{-0.026}^{+0.029}$}	& \multicolumn{2}{c}{$-0.300_{-0.043}^{+0.044}$} \\[3pt]
		      e				& $0.18\pm0.02$ 	& \multicolumn{2}{c}{$0.146_{-0.018}^{+0.018}$}		& \multicolumn{2}{c}{$0.224_{-0.014}^{+0.014}$} \\[3pt]
		      $\omega$ (deg)		& $232.5\pm9.4$ 	& \multicolumn{2}{c}{$259.217_{-6.744}^{+6.475}$}	& \multicolumn{2}{c}{$219.541_{-6.591}^{+6.821}$} \\[3pt]
		      M$\sin{i}$ (M$_J$)	& $1.90\pm0.06$ 	& \multicolumn{2}{c}{$1.573_{-0.026}^{+0.026}$}		& \multicolumn{2}{c}{$1.985_{-0.033}^{+0.033}$} \\[3pt]
		      a (AU)			& $2.07\pm0.06$ 	& \multicolumn{2}{c}{$1.836_{-0.016}^{+0.015}$}		& \multicolumn{2}{c}{$2.231_{-0.019}^{+0.019}$} \\[3pt]
		      T$_{peri}$ (days)		& $2453869\pm24$ 	& \multicolumn{2}{c}{$2456552.8_{-21.0}^{+16.9}$}	& \multicolumn{2}{c}{$2457373.0_{-24.5}^{+25.2}$} \\[3pt]
		    \hline
			n$_{obs}$		& 20			& \multicolumn{4}{c}{44} \\[3pt]
		    \hline
						& mean RV error	& $\gamma$		& jitter	& $\gamma$		& jitter\\
						& (ms$^{-1}$)	& (ms$^{-1}$)		& (ms$^{-1}$)	& (ms$^{-1}$)		& (ms$^{-1}$)\\[3pt]
		    \hline
		      HARPS			& 0.65		& $-11.78_{-0.36}^{+0.37}$&$1.32_{-0.18}^{+0.22}$& $-2.62_{-0.57}^{+0.61}$& $1.39_{-0.17}^{+0.21}$ \\[3pt]
		    \hline
		      BIC			&		& \multicolumn{2}{c}{$99.79$}			& \multicolumn{2}{c}{$101.21$} \\[3pt]
		    \hline
		  \end{tabular}
	  \end{table*}
	  
	\paragraph{HD 216437}
	
	  A G2IV star whose planetary companion was independently detected by \citet{jones2002} using 39 UCLES 
	  datapoints and by \citet{mayor2004} with 21 CORALIE observations, finding similar solutions.
	  \par
\begin{figure}
	    \includegraphics[width=\linewidth]{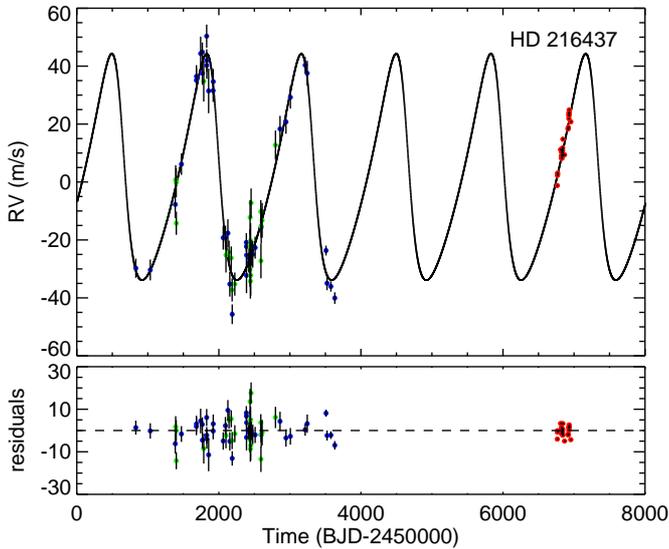}
	    \caption{Same as Fig. \ref{fig:hd4208-fit} but for system HD 216437. The literature UCLES data are shown in blue, CORALIE	
		      datapoints are in green while our HARPS survey data are shown in red.}
	    \label{fig:hd216437-fit}
	  \end{figure}

 Using the 33 HARPS datapoint obtained from our observations we find (see Table \ref{table:hd216437} and Fig. 
	  \ref{fig:hd216437-fit}) significantly 
	  higher values for minimum mass M$\sin{i}$=$2.223_{-0.058}^{+0.058}$M$_J$ and period P=$1334.28_{-13.36}^{+13.07}$d, while 
	  the resulting eccentricity value of e=$0.317_{-0.027}^{+0.028}$ is consistent with the published solution but has a higher 
	  accuracy. Hill's criterion gives an orbital period of 427.1 days as a limit on an inner planet's dynamical stability.
	  \par No low FAP peak is found in the residual data periodogram.	  
	  
	  \begin{table} 
	    \small
		  \caption{Fit results comparison for system HD 216437}\label{table:hd216437}
		  \centering
		  \begin{tabular}{l c c c}
		    \hline\hline
		      \multicolumn{4}{c}{HD 216437}\\
		    \hline
						& \multicolumn{2}{c}{\cite{mayor2004}}	& This work\\
		      Parameter			& \multicolumn{2}{c}{Planet b}		& Planet b\\
		    \hline
		      K (ms$^{-1}$)		& \multicolumn{2}{c}{$34.6\pm5.7$} 	& $39.08_{-1.04}^{+1.07}$ \\[3pt]
		      P (days)			& \multicolumn{2}{c}{$1256\pm35$} 	& $1334.28_{-13.36}^{+13.07}$ \\[3pt]
		      $\sqrt{e}\cos{\omega}$	& \multicolumn{2}{c}{-}			& $0.237_{-0.052}^{+0.051}$ \\[3pt]
		      $\sqrt{e}\sin{\omega}$	& \multicolumn{2}{c}{-}			& $0.509_{-0.028}^{+0.026}$ \\[3pt]
		      e				& \multicolumn{2}{c}{$0.29\pm0.12$} 	& $0.317_{-0.027}^{+0.028}$ \\[3pt]
		      $\omega$ (deg)		& \multicolumn{2}{c}{$63\pm22$} 	& $65.092_{-5.394}^{+5.515}$ \\[3pt]
		      M$\sin{i}$ (M$_J$)	& \multicolumn{2}{c}{$1.82$} 		& $2.223_{-0.058}^{+0.058}$ \\[3pt]
		      a (AU)			& \multicolumn{2}{c}{$2.32$} 		& $2.497_{-0.037}^{+0.036}$ \\[3pt]
		      T$_{peri}$ (days)		& \multicolumn{2}{c}{$2450693\pm130$} 	& $2457289.2_{-46.6}^{+52.2}$ \\[3pt]
		    \hline
			n$_{obs}$		& \multicolumn{2}{c}{21}		& 93 \\[3pt]
		    \hline
						& mean RV error	& $\gamma$		& jitter	\\
						& (ms$^{-1}$)	& (ms$^{-1}$)		& (ms$^{-1}$)\\[3pt]
		    \hline
		      UCLES			& 3.86		& $-0.05_{-1.02}^{+1.01}$ & $4.11_{-0.75}^{+0.912}$ \\[3pt]
		      CORALIE			& 5.33		& $16.13_{-1.94}^{+1.90}$ & $6.47_{-1.65}^{+1.96}$ \\[3pt]
		      HARPS			& 0.32		& $-12.41_{-5.36}^{+5.06}$ & $2.07_{-0.25}^{+0.30}$ \\[3pt]
		    \hline
		  \end{tabular}
	  \end{table}
	  
	\paragraph{HD 220689}
	
	  This G3V star's planetary companion discovery was reported in \citet{marmier2013} after the acquisition 
	  of 48 CORALIE datapoints, the last 34 of which were obtained after an instrument upgrade and are therefore treated as an 
	  independent dataset. This planet is also notable for having the lowest radial velocity semiamplitude detected by CORALIE at 
	  the time of publication.
	  \par
\begin{figure}
	    \includegraphics[width=\linewidth]{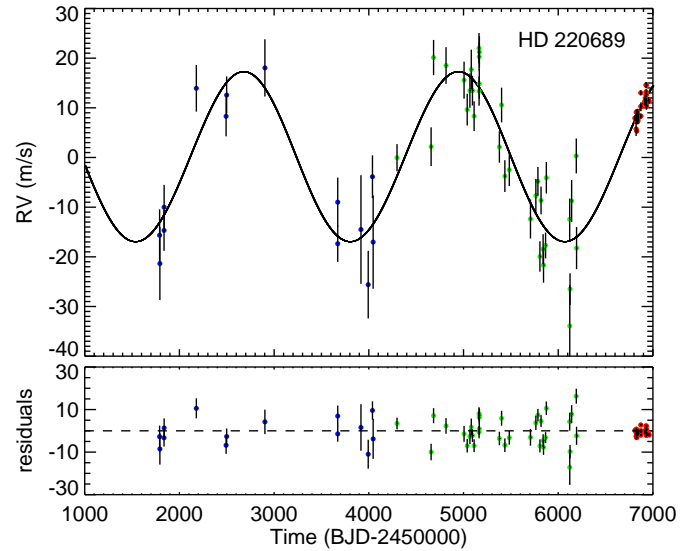}
	    \caption{Same as Fig. \ref{fig:hd4208-fit} but for system HD 220689. The literature CORALIE surveys data are shown in blue and 
		      green, while our HARPS survey data are shown in red.}
	    \label{fig:hd220689-fit}
	  \end{figure}

 Using the 31 HARPS datapoints we collected, we found  a compatible but better constrained value for orbital period 
	  P=$2266.40_{-58.77}^{+65.84}$d, a higher precision value of eccentricity e=$0.054_{-0.038}^{+0.061}$ and a similarly better 
	  determined value of minimum mass M$\sin{i}$=$1.118_{-0.035}^{+0.035}$M$_J$ (see Table \ref{table:hd220689} and Fig. 
	  \ref{fig:hd220689-fit}). We find the maximum period allowing dynamical stability for an additional inner planet to be 1426.6 
	  days. No peak under 1\% level of false alarm probability was found in the residual periodogram.	  
	
	  \begin{table} 
	    \small
		  \caption{Fit results comparison for system HD 220689}\label{table:hd220689}
		  \centering
		  \begin{tabular}{l c c c}
		    \hline\hline
		      \multicolumn{4}{c}{HD 220689}\\
		    \hline
						& \multicolumn{2}{c}{\cite{marmier2013}}	& This work\\
		      Parameter			& \multicolumn{2}{c}{Planet b}			& Planet b\\
		    \hline
		      K (ms$^{-1}$)		& \multicolumn{2}{c}{$16.4\pm1.5$}		& $17.12_{-1.19}^{+1.26}$ \\[3pt]
		      P (days)			& \multicolumn{2}{c}{$2209_{-81}^{+103}$}	& $2266.40_{-58.77}^{+65.84}$ \\[3pt]
		      $\sqrt{e}\cos{\omega}$	& \multicolumn{2}{c}{-} 			& $0.090_{-0.185}^{+0.161}$ \\[3pt]
		      $\sqrt{e}\sin{\omega}$	& \multicolumn{2}{c}{-} 			& $0.059_{-0.193}^{+0.176}$ \\[3pt]
		      e				& \multicolumn{2}{c}{$0.16_{-0.07}^{+0.10}$}	& $0.054_{-0.038}^{+0.061}$ \\[3pt]
		      $\omega$ (deg)		& \multicolumn{2}{c}{$137\pm75$} 		& $112.834_{-80.118}^{+194.3244}$ \\[3pt]
		      M$\sin{i}$ (M$_J$)	& \multicolumn{2}{c}{$1.06\pm0.09$} 		& $1.118_{-0.035}^{+0.035}$ \\[3pt]
		      a (AU)			& \multicolumn{2}{c}{$3.36\pm0.09$} 		& $3.396_{-0.081}^{+0.084}$ \\[3pt]
		      T$_{peri}$ (days)		& \multicolumn{2}{c}{$245649\pm413$} 		& $2457532.9_{-478.1}^{+772.5}$ \\[3pt]
		    \hline
			n$_{obs}$		& \multicolumn{2}{c}{48}			& 79 \\[3pt]
		    \hline
						& mean RV error	& $\gamma$		& jitter	\\
						& (ms$^{-1}$)	& (ms$^{-1}$)		& (ms$^{-1}$)\\[3pt]
		    \hline
		      CORALIE1			& 5.67		& $6.87_{-2.51}^{+2.40}$ & $4.85_{-2.25}^{+2.46}$ \\[3pt]
		      CORALIE2			& 3.48		& $-0.07_{-1.39}^{+1.38}$& $6.02_{-0.95}^{+1.16}$ \\[3pt]
		      HARPS			& 0.59		& $-9.89_{-2.66}^{+3.16}$& $1.35_{-0.19}^{+0.23}$ \\[3pt]
		    \hline
		  \end{tabular}
	  \end{table}
    
  \section{Detection limits and planetary frequency}	\label{sec:detectionlimits}
    
    The calculation of the planetary detection frequency in the star sample was carried out using a standard procedure, which is the 
    production of 
    synthetic circular radial velocity curves for different realizations of the orbital elements pair (P, M$\sin{i}$) and the 
    statistical analysis of the obtained RV curve in order to determine the detection probability of the signal produced by the 
    simulated planet; a recent application of this technique can be seen in \citet{faria2016}.
    \par For each star in our sample we explored 100 different orbital periods $P_{j}$ evenly spaced in logarithm from a single day 
    to one year and 100 different planetary mass $M_{k}$ similarly spaced from 1 M$_\oplus$ to 2 M$_J$. For each one of these 
    period-mass realization we then computed 20 synthetic radial velocity curves, each one with a different randomly obtained 
    combination of (T$_{peri}$, $\omega$); these synthetic radial velocities were obtained by evaluating at each of our survey's 
    observation times $t_i$ the theoretical radial velocity value RV($t_i,P_{j},M_{k},T_{peri},\omega$) for the injected period and 
    mass, and then adding to this theoretical values a random gaussian noise with an amplitude equal to the standard deviation of the 
    post-fit 
    residuals data for our high cadence and precision HARPS survey. We then produce a periodogram for each one of the 2$\cdot10^5$ 
    synthetic signals thus obtained, and the signal is considered as detected only if the injected period power is higher than the 
    power corresponding to a 1\% false alarm probability.
    \par It is important to note that our choice to use only our high precision and cadence HARPS residual data for this synthetic 
    curves production leads to a rather conservative evaluation of the detection frequency, being the typical RMS of our post-fit 
    residual data around 3-4 ms$^{-1}$ rather than the $\sim0.5$ ms$^{-1}$ RMS of our original HARPS data. Also, choosing not to 
    include the lower-precision literature data could lead to some precision loss at high periods, a fact that has however low influence 
    on our analysis since we choose to study detection for periods up to 1 year. It is also important to note that only in three of 
    the systems of our sample (HD 23127, HD 38801 and HD 48265) a limited portion of the outer region of the period range being 
    considered in this detection 
    limits analysis is subject to instability as calculated throught the Hill criterion (see Sect. \ref{sec:analysis} and each 
    system's paragraph in Subsect. \ref{subsec:casebycase}).
  
      \begin{figure}
	\includegraphics[width=\linewidth]{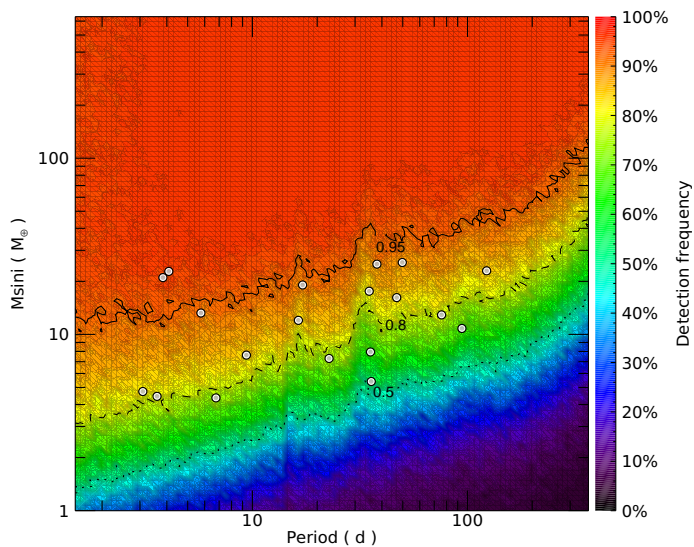}
	\caption{Color-coded HARPS precision detection frequency map for the whole 20-systems sample studied in this paper, period 
		  ranging from one day to one year and masses ranging from 1 M$_\oplus$ to 2 M$_J$. The detection frequency levels of 
		  50\%, 80\% and 95\% are respectively shown as dotted, dashed and solid curves. The low-mass inner planets of 
		  the 11 archival Solar System analogs discussed in Sect. \ref{sec:introduction} are shown as white circles. }
	\label{fig:survey-det}
      \end{figure}
      
    \par The detection map produced for the overall HARPS survey, obtained summing and renormalizing each and every 
    system's detection map, is shown in Fig. \ref{fig:survey-det}. It can be seen that true uniform completeness for the whole 20-stars 
    sample is achieved for periods below 50 days only for M$\sin{i}$>30 M$_\oplus$, while for periods less than 150 days we have whole 
    sample completeness only for M$\sin{i}$>50 M$_\oplus$. For the same period limits we are instead uniformely sensitive for half our 
    sample respectively for minimum masses ranging from 5 to 10 Earth masses and from 7 to 15 Earth masses. We also note that 
    that for minimum masses ranging from 10 to 30 M$\oplus$ we have completeness below 10 days, a fact that in combination with the 
    lack of additional hot Neptunes found in our analysis suggest a strong absence of such planets around the selected stars.
    \par It is therefore clear that to achieve higher detection frequencies for terrestrial and super-terrestrial planet at all periods, 
    expecially shorter ones, it is necessary to obtain a higher number of observations per target star and ensure a higher-density 
    sampling. Our observing program, as noted in Sect. \ref{sec:observations}, originally foresaw about 40 measurements per star, a 
    much 
    higher number than the average 27 per star we actually obtained; it can be argued that a number of observations closer to the one 
    originally proposed could push the survey's detection limits to lower planetary masses, ensuring a better coverage and higher 
    detection frequencies for terrestrial and super-terrestrial additional planets. In fact, analytical expressions for 
    close-in low-mass exoplanets detection threshold found in \cite{narayan2005} show such a dependence on the number of radial 
    velocity measurements.
    It will also be important in the near future  
    comparing our results with the similarly produced detection map for the parallel HARPS-N survey mentioned in 
    Sect. \ref{sec:introduction}.
    \par From the overall survey detection map one can finally obtain an estimate for the planetary occurence frequency $f_p$ in the 
    presence of a giant outer planet; the probability of obtaining $m$ detection out of a sample of size $N$ is given by the binomial 
    distribution:
      $$ p(m;N,f_p)=\frac{N!}{m!(N-m)!}\ f_p^m (1-f_p)^{N-m} $$
    The choice of distribution is dictated by the small size of our sample (at most $N$=20) and has been originally justified in 
    \citet{burgasser2003} and successfully used in \citet{sozzetti2009} and \citet{faria2016}, to name a few. 
    \par During the analysis of our sample, as detailed in Sect. \ref{sec:analysis}, we found a single new candidate planet, 
    namely the gas giant HD 50499 c, which due to its estimated minimum orbital period P$\geq$8265.51 days lies outside our 
    considered period range of P$\leq$1 yr. Following a section of the analysis reported in the work of \citet{mayor2011}, in 
    which the authors estimated the 
    occurrence frequency of stars with at leats one planet in given regions of orbital period and planetary mass, we focus our 
    interest on sub-giant planets with orbital period less than 150 days and M$\sin{i}$ between 10 and 30 M$_\oplus$, a region 
    for which we are conservatively sensitive to only 50\% of our 20-stars sample. Having found no 
    candidate planet falling within this mass-period space, we can only obtain a 1-$\sigma$ lower limit for occurence 
    frequency of low-mass planets in the 
    presence of a known outer giant planet of $f_p$<9.84\%, a value we note to be much lower than the 38.8$\pm$7.1\% reported in 
    \citet{mayor2011} for stars hosting at least one planet in the same period-mass region.
  
  \section{Summary and discussion}	\label{sec:discussion}
    In this work we have reported the results of an intense observational campaign conducted using the high-precision HARPS 
    spectrograph on 20 stars, obtaining an average of 27 datapoints per star. The stars in the sample were selected in virtue of 
    being bright, inactive and not significantly evolved Sun-like stars hosting at the time of selection a single long-period giant 
    planet previously discovered via radial velocity observations, the final objective of our observations being the search for 
    additional inner low-mass planets and estimating the occurrence frequency of scaled-down Solar System analogs.
    \par By joining the literature radial velocity data with our own measurements we have obtained revised orbital solutions for 
    each known planet in our sample using an MCMC-based fitting algorithm, generally characterized by a higher precision on most of 
    the orbital parameters and significant updates on the orbital 
    parameters for half of the selected systems. We especially stress the achievement of a drastically different set of orbital elements 
    for the previously characterized planet HD 30177 c and the characterization of previously unpublished outer giant planet HD 50499 c, 
    both results obtained fitting the data with a Keplerian curve with a parabolic trend.
    \par Also, three of the systems in our sample (HD 66428, HD 73267 and HD 152079) show a significant slope in 
    the data, one of which (HD 73267) was previously unpublished, suggesting the existence of at least one additional outer companion 
    in the system having a minimum mass >0.83 M$_J$.
    \par We have also conducted detection simulations on all sample stars in order to calculate the occurrence rate of inner 
    (periods from one day to one year) low mass (10-30 M$_\oplus$) planets in the presence of long-period giants. Having found no 
    candidate planet within these period and mass ranges we can only provide an estimate for the upper limit of said frequency, namely
    $f_p$<$9.84\%$, a value which is significantly lower than the one obtained for the same period-mass ranges by \citet{mayor2011} 
    for stars hosting at least one planet.
    \par The lack of candidate planets found on inner orbits in the sample systems is especially significant when considering again the 
    detection limits of our sample, shown in Fig. \ref{fig:survey-det}, from which is clear that we are sensitive to the vast majority 
    of "hot super-Earths" or "mini-Neptunes", defined as planets up to 20 times more massive than the Earth on orbits shorter than 100 
    days. This type of planet is found to be present around roughly half of all Sun-like stars 
    \citep{mayor2011, howard2012, izidoro2015, morbidelli2016}, often in a compact low-eccentricity configuration within multiple 
    systems, and is noticeably missing in our own Solar System. The search for the reason of the lack of hot super-Earth in the 
    Solar System has produced several competing formation models for super-Earths \citep[see][]{morbidelli2016}, roughly 
    distinguishable into in situ formation \citep{hansenmurray2012,hansenmurray2013,martinlivio2016} and inward migration processes 
    \citep{cossou2014,izidoro2015,izidoro2017}. The latter model, proposing that super-Earth embryos form in the outer protoplanetary 
    disk before migrating inward, is especially interesting in interpreting the lack of inner candidate planets found in our sample 
    because, as detailed in \cite{izidoro2015}, if the innermost super-Earth embryo grows into a giant planet core while migrating 
    inward the newly produced gas giant can and will significantly alter the dynamical evolution of the planetary system, blocking the 
    outer super-Earths from further inward migration, with some embryos only occasionally crossing the giant planet's orbit and 
    becoming hot super-Earths or mini-Neptunes. While some studies 
    \citep{volk2015,batygin2015} propose that the early 
    Solar System hosted a close-in population of super-Earths that was either destroyed by erosive collisions or by merging with the 
    Sun, \cite{izidoro2018} points out that the debris generated by erosive processes should have catalyzed further growth of close-in 
    planets and that the existence of an inner edge in the protoplanetary disk should actually prevent planets from simply falling 
    onto the Sun, suggesting instead that the wide, low-eccentricity orbit of Jupiter has had a key role in effectively blocking 
    the formation of close-in super-Earths in the Solar System.
    The lack of planets found on inner orbits in our sample could be interpreted as 
    evidence for the main observational prediction of the inward migration model, namely the anti-correlation between planetary systems 
    hosting close-in super-Earths and those hosting long-period gas giants.
    \par There is also an interesting formal match between the occurence rate $9.84\%$ of low mass planets in the presence of 
    outer giants derived from our analysis and the preliminary $\sim10\%$ fraction of Solar System analogs discussed in Sect.
    \ref{sec:introduction} that we found amongst the multi-planet systems studied via the radial velocity method; it is also 
    interesting to note from \ref{fig:survey-det} that the inner low-mass planets in these 11 systems would have been detected by our 
    survey in at least 50\% of the cases. It could be also argued that the low number of low-mass inner planets in these systems 
    could be examples of the planetary embryos that can occasionally cross the innermost giant planet's orbit in the inward migration 
    formation scenario. However, we stress that the 11 multi-planet systems are not a homogeneous sample, showing a variety of stellar 
    type and planetary characteristics and that were selected in our preliminary search in virtue of the relation between planetary 
    orbits in relation and the circumstellar habitable zone, while the 20-stars sample that is the main focus of this work were all 
    homogeneously selected 
    for their stellar characteristics (as detailed in Sect. \ref{sec:sample}) and that we did not focus on their habitable zone. 
    While it 
    is interesting to note a suppression of this type of planetary architecture across different spectral types, directly 
    comparing such differently selected groups of systems to draw supportable conclusions about their dynamical history is  
    difficult without further data and higher statistics and beyond the scope of this work.
    \par We finally stress the need for further high-cadence and high-precision observations on the sample stars in order to better 
    characterize certain systems; in particular the constrains on some planets eccentricity and longitude of periastron, such as 
    HD 38801 b and HD 126525 b will greatly benefit from additional 
    intense observations. Exceptional attention will surely be needed in future observations of systems HD 30177 and HD 50499, whose 
    long-period (respectively $\simeq$17 and $\simeq$22 years) outer planet orbits are still incompletely sampled and therefore only 
    partially characterized via semiamplitude, period and mass lower limits.
    \par As a final and future-oriented point of interest we estimated the astrometric signal $\alpha$ of each known giant planet 
    in our sample, using the relation:
      $$ \alpha=\frac{M_{p}}{M_*}\frac{a}{d} $$
    in order to obtain a quick evaluation of how many and which systems in our sample could be observed by the high-precision 
    astrometric satellite Gaia during its ongoing five-year mission, to fuel further instrument synergies. We find 13 of our systems 
    (namely HD 27631, HD 30177, HD 38801, HD 50499, HD 66428, HD 70642, HD 73267, HD 117207, HD 143361, HD 152079, HD 190647, 
    HD 216437 and HD 220689) to have an astrometric signal $\alpha\gtrsim50$ $\mu$as (expected to be detected by Gaia at high S/N); 
    being however $\alpha$ 
    obviously dependent on the planet's true mass this fraction of astrometrically observable planets could very well be much higher if 
    their orbital inclination turns out to be rather small in value. The importance of evaluating which systems could be observed by 
    Gaia is clear when considering that during its five-year mission the satellite will provide a partial or complete coverage of the 
    known planet's orbit, producing therefore a high-accuracy dataset that could be joined with the literature radial velocity data 
    to further constrain the systems orbital elements and search for additional inner companions in the residuals, an important 
    example of how using different detection methods and instruments can and will help future detailed analysis on exoplanetary systems 
    architecture.
    
  \begin{acknowledgements}
    We thank the anonymous referee for useful comments.
    DB acknowledges financial support from INAF and Agenzia Spaziale Italiana (ASI grant n. 014-025-R.1.2015) for the 2016 PhD 
    fellowship programme of INAF. 
    MD acknowledges funding from INAF through the Progetti Premiali funding scheme of the Italian Ministry of Education, University, 
    and Research. 
    The research leading to these results has received funding from the European Union Seventh Framework Programme (FP7/2007-2013) 
    under Grant Agreement No. 313014 (ETAEARTH).
  \end{acknowledgements}
  
  \bibliographystyle{aa}
  \bibliography{ref}
  \clearpage
    
\longtab{
\begin{longtable}{l c c c c c c c c}
	\caption{RV velocity tables}\label{table:data}\\
	\hline\hline
	BJD & T$_{exp}$ & RV  & BIS & FWHM & Ca II  & H$\alpha$  & Air mass & SNR \\
	    & [$\mathrm{s}$] & [$\mathrm{kms^{-1}}$] & [$\mathrm{kms^{-1}}$]&[$\mathrm{kms^{-1}}$] & & & & \\
	\hline
	\endfirsthead
	\caption{continued.}\\
	\hline
	BJD & T$_{exp}$ & RV  & BIS & FWHM & Ca II  & H$\alpha$  & Air mass & SNR \\
	    & [$\mathrm{s}$] & [$\mathrm{kms^{-1}}$] & [$\mathrm{kms^{-1}}$]&[$\mathrm{kms^{-1}}$] & & & & \\
	\hline
	\endhead
	\hline
	\endfoot

	\multicolumn{9}{c}{HD 4208}\\
	\hline
		2456810.929653 & 900 & 56.8184 $\pm$ 0.0005 & -0.037 & 6.564 & 0.00343 $\pm$ 0.00004 & 0.03098 $\pm$ 0.00006 & 1.20 & 113.10\\
		2456823.914287 & 900 & 56.8150 $\pm$ 0.0005 & -0.037 & 6.561 & 0.00345 $\pm$ 0.00004 & 0.03097 $\pm$ 0.00005 & 1.13 & 119.65\\
		2456824.882368 & 900 & 56.8158 $\pm$ 0.0004 & -0.037 & 6.565 & 0.00343 $\pm$ 0.00004 & 0.03107 $\pm$ 0.00005 & 1.25 & 121.50\\
		2456824.894267 & 900 & 56.8170 $\pm$ 0.0004 & -0.038 & 6.566 & 0.00344 $\pm$ 0.00003 & 0.03092 $\pm$ 0.00005 & 1.19 & 129.65\\
		2456825.936023 & 900 & 56.8187 $\pm$ 0.0004 & -0.041 & 6.565 & 0.00349 $\pm$ 0.00003 & 0.03098 $\pm$ 0.00005 & 1.06 & 127.90\\
		2456826.899682 & 900 & 56.8173 $\pm$ 0.0004 & -0.038 & 6.566 & 0.00342 $\pm$ 0.00003 & 0.03092 $\pm$ 0.00004 & 1.15 & 150.05\\
		2456826.945334 & 900 & 56.8166 $\pm$ 0.0004 & -0.039 & 6.565 & 0.00346 $\pm$ 0.00003 & 0.03098 $\pm$ 0.00004 & 1.04 & 145.65\\
\hline
	\end{longtable}
	}

\end{document}